\documentclass[fleqn,usenatbib]{mnras}

\usepackage{newtxtext,newtxmath}
\usepackage[T1]{fontenc}

\DeclareRobustCommand{\VAN}[3]{#2}
\let\VANthebibliography\thebibliography
\def\thebibliography{\DeclareRobustCommand{\VAN}[3]{##3}\VANthebibliography}

\usepackage{graphicx}	% Including figure files
\usepackage{amsmath}	% Advanced maths commands

\title[Simultaneously Modelling DSFGs and MQs]{Simultaneously Modelling Dusty Star Forming Galaxies and Massive Quiescents: A Calibration Framework for Galaxy Formation Models}

\author[P. Araya-Araya et al.]{
Pablo Araya-Araya,$^{1}$\thanks{E-mail: paraya-araya@usp.br}
Rachel K. Cochrane,$^{2,3,4}$\thanks{E-mail: rcochra3@roe.ac.uk}
Christopher C. Hayward,$^{5,6}$
Laerte Sodré Jr.,$^{1}$ 
\newauthor 
\ Robert M. Yates,$^{7}$
Marcel P. van Daalen,$^{8}$
and Marcelo C. Vicentin$^{1}$
\\
$^{1}$Departamento de Astronomia, Instituto de Astronomia, Geofísica e Ciências Atmosféricas,
Universidade de São Paulo, 
Rua do Matão 1226, \\ Cidade Universitária, 05508-900, São Paulo, SP, Brazil\\
$^{2}$Institute for Astronomy, University of Edinburgh, Royal Observatory, Blackford Hill, Edinburgh, EH9 3HJ, UK\\
$^{3}$Columbia Astrophysics Laboratory, Columbia University, 550 West 120th Street, New York, NY 10027, USA\\
$^{4}$Center for Computational Astrophysics, Flatiron Institute, 162 Fifth Avenue, New York, NY 10010, USA\\
$^{5}$Eureka Scientific, Inc., 2452 Delmer Street, Suite 100, Oakland, CA 94602, USA\\
$^{6}$Kavli Institute for the Physics and Mathematics of the Universe (WPI), The University of Tokyo Institutes for Advanced Study, \\ The University of Tokyo, Kashiwa, Chiba 277-8583, Japan\\
$^{7}$Centre for Astrophysics Research, University of Hertfordshire, Hatfield, AL10 9AB, UK\\
$^{8}$Leiden Observatory, Leiden University, PO Box 9513, NL-2300 RA Leiden, the Netherlands
}

\date{Accepted XXX. Received YYY; in original form ZZZ}

\pubyear{\the\year{}}

\begin{document}
\label{firstpage}
\pagerange{\pageref{firstpage}--\pageref{lastpage}}
\maketitle

\begin{abstract}
Galaxy formation models, particularly semi-analytic models (SAMs), rely on differential equations with free parameters to describe the physical mechanisms governing galaxy formation and evolution. Traditionally, most SAMs calibrate these parameters manually to match observational data. However, this approach fails to fully explore the multidimensional parameter space, resulting in limited robustness and inconsistency with some observations. In contrast, the \texttt{L-Galaxies} SAM features a unique Markov Chain Monte Carlo (MCMC) mode, enabling robust model calibration. Using this functionality, we address a long-standing tension in galaxy formation models: simultaneously reproducing the number densities of dusty star-forming galaxies (DSFGs) and high-redshift massive quiescent galaxies (MQs). We test nine combinations of observational constraints — including stellar mass functions, quiescent fractions, neutral hydrogen mass functions, and DSFG number densities — across different redshifts. We then analyze the resulting galaxy property predictions and discuss the underlying physical mechanisms. Our results identify a model that reasonably matches the number density of DSFGs while remaining consistent with observationally-derived lower limits on the number density of high-redshift MQs. This model requires high star formation efficiencies in mergers and a null dependency of supermassive black hole (SMBH) cold gas accretion on halo mass, facilitating rapid stellar mass and SMBH growth. Additionally, our findings highlight the importance of robust calibration procedures to address the significant degeneracies inherent to multidimensional galaxy formation models.
\end{abstract}

\begin{keywords}
methods: numerical – galaxies: evolution – galaxies: formation – galaxies: high-redshift
\end{keywords}

%%%%%%%%%%%%%%%%%%%%%%%%%%%%%%%%%%%%%%%%%%%%%%%%%%

%%%%%%%%%%%%%%%%% BODY OF PAPER %%%%%%%%%%%%%%%%%%

\section{Introduction} \label{sec:intro}
Dusty star-forming galaxies (DSFGs, also known as sub-millimetre galaxies or SMGs) have gained significant attention since their discovery in the late 1990s \citep{smail97, barger98, hughes98, eales99}. Intrinsically, DSFGs are highly luminous ($L_{\rm IR} \gtrsim 10^{11}\,\rm{L_{\odot}}$), and due to the negative $k$-correction, they are relatively easy to detect even at high redshifts \citep[e.g. at $z\gtrsim4$;][]{Cooper2022,Manning2022,Long2024}. DSFGs were first identified in single-dish surveys, where large beam sizes hampered individual source localisation and cross-matching to multi-wavelength data \citep[e.g. see the extensive work that led to a secure redshift for the SMG HDF850.1;][]{hughes98,Downes99,Richards99,Dunlop04,Wagg07,Cowie09,Walter12,Neri14,Herard-Demanche25}. \\
\indent The high angular resolution of the Atacama Large Millimeter/sub-millimetre Array (ALMA) has enabled the localisation and detailed characterization of DSFGs across cosmic time \citep[see][for a review]{hodge20}, leading to constraints on their redshift distribution, physical properties and large-scale environments. Observational studies show that DSFGs are predominantly found at redshifts $z \sim 2-3$ \citep{chapman05, simpson17, dud20}, approximately coinciding with the global peak of cosmic star formation activity \citep[e.g.][]{madau14,rachel23b}. Dust-obscured star formation comprises nearly half of the total cosmic star formation rate density (CSFRD) at these epochs \citep{dunlop17, michalowski17, smith17, zavala21}. Despite this, DSFGs are relatively rare, with number densities of $N \sim 10^{-5}\,\rm{Mpc}^{- 3}\,\rm{Gyr}^{-1}$ at the peak of their redshift distribution \citep{dud20}. The brightest DSFGs are highly clustered \citep{Blain2004,Chen2016,Garcia2020,Lim2020,Stach21}, and also serve as effective tracers of galaxy protoclusters \citep{chapman01, daddi09, dannerbauer14, casey16, Marrone18, miller18, oteo18, wang21, gouin22, calvi23, yo24, hill24,Herard-Demanche25}. They have been proposed as potential progenitors of the massive elliptical galaxies found at the centers of present-day galaxy clusters \citep[e.g.][]{toft14}. Their extreme properties — including high stellar masses ($M_{\star} \sim 10^{11}\,\rm{M_{\odot}}$) and intense dust-obscured star formation rates (SFR $\sim 10^2-10^3\,\rm{M_{\odot}/yr}$) \citep[e.g.][]{Simpson2014,dacunha15,dud20,Cochrane21} — make DSFGs valuable laboratories for both observational and theoretical studies of galaxy evolution in extreme environments.\\
\indent Historically, theoretical models have struggled to reproduce the DSFG population \citep[see Section 10 of ][for a review]{casey14}, particularly sub-millimetre (sub-mm) number counts (e.g. at $870\,\mu\rm{m}$; \citealt{granato00, fontanot07, somerville12, cowley19, chris21}). A potential solution was proposed by \citet{baugh05}, who used the \texttt{GALFORM} semi-analytic model (SAM) \citep{cole00, lacey16} to show that a top-heavy\footnote{a d$n$/d$\log m$ constant is assumed in \citet{baugh05} while a d$n$/d$\log m \propto m^{-1}$ is implemented by \citet{lacey16}.} stellar initial mass function (IMF) in merger-induced starbursts could bring models into better agreement with observations. However, this modification was controversial, as SAMs (and galaxy formation models in general) contain multiple free parameters, allowing for alternative solutions that do not require an IMF variation. Indeed, \citet{chris13} demonstrated this by statistically estimating sub-mm number counts under the assumption of a universal IMF, calculating flux densities using relations between dust mass, star formation rate and sub-mm flux density derived from high-resolution radiative transfer calculations \citep{chris11}. This conclusion was reinforced by \citet{safarzadeh17}, who applied the \citet{lu11, lu14} SAM in combination with the scaling relations derived by \citet{chris13}. Similarly, \citet{lagos19} found agreement with observed sub-mm number counts using the \texttt{SHARK} SAM \citep{lagos18}, while also assuming a universal IMF.\\
\indent The success of large-box cosmological hydrodynamical simulations in reproducing the statistical properties of SMGs has been mixed. Using sub-mm flux densities calculated via radiative transfer post-processing on galaxies drawn from the largest-box EAGLE simulation \citep{McAlpine2016,trayford17}, \cite{mcalpine19} explored the nature of simulated galaxies with $850\,\mu\rm{m}$ flux density above $1\,\rm{mJy}$. Although this sample broadly reproduced the redshift distribution of observed SMGs, number counts of the most extreme SMGs were underpredicted by over an order of magnitude \citep{cowley19}. Deriving sub-mm flux densities from scaling relations based on dust mass and SFR, \citet{chris21} showed that the Illustris simulation \citep{Genel2014,vogelsberger14,Sijacki2015} better matches sub-mm number counts, but its successor IllustrisTNG \citep{Marinacci2018,Springel2018,Nelson2018a,pillepich18,Naiman2018} fails to do so. This seemed to be reflective of the lower dust masses and star formation rates seen for high-mass galaxies in IllustrisTNG, where quenching happens earlier (the redshift distribution of bright SMGs also peaks at higher redshift than is observed; \citealt{kumar25}). \citet{lovell21} demonstrated that the \texttt{SIMBA} simulation \citep{dave19}, when coupled with the radiative transfer code \texttt{POWDERDAY} \citep{narayanan21}, produces number counts consistent with observations; nevertheless, the redshift distribution of their brightest sources appears skewed to higher redshifts than is inferred from observations. More recently, \citet{kumar25} incorporated scaling relations into the \texttt{FLAMINGO} simulation \citep{schaye23, kugel23}, successfully reproducing observed sub-mm number counts when the \cite{chris13} $S_{870\,\mu\rm{m}}$ calibration is applied (though still underpredicting these using the updated \citealt{lovell21} calibration) and also matching the observationally-inferred redshift distribution.\\
\indent As with the sub-mm number counts, predicting the observed number density of massive quiescent galaxies (MQs; $M_{\star} \gtrsim 10^{10.5}\,\rm{M_{\odot}}$ and sSFR $\lesssim 10^{-11}$ yr$^{-1}$) at high redshift has been a major challenge for theoretical models, particularly in light of recent JWST observations. For instance, \citet{lagos25} compared how well different models — \texttt{GAEA} \citep{delucia24}, \texttt{GALFORM}, \texttt{SHARK} (semi-analytic models, SAMs), as well as \texttt{SIMBA}, \texttt{IllustrisTNG}, and \texttt{EAGLE} (hydrodynamical simulations) — reproduce the number density of MQs. Their analysis found that all these simulations underpredict MQs by between $0.3\,\rm{dex}$ and several dex when compared to recent JWST results \citep{valentino23, carnall23, nanayakkara24, alberts24}. Similar discrepancies were reported by \citet{szpila25} using the \texttt{SIMBA}-C simulation \citep{hough23, hough24} and by \citet{vani25} with the \citet{ayromlou21} version of \texttt{L-Galaxies} SAM. On the other hand, \citet{kimmig25} reported good agreement between the \texttt{MAGNETICUM} Pathfinder simulation \citep{Hirschmann2014,Ragagnin2017} and results from JWST observations regarding the MQ population at high redshift. However, this model significantly overpredicts the MQ population (by an order of magnitude) at lower redshifts \citep{lagos25}.\\
\indent Overall, while some simulations successfully reproduce the sub-mm number counts, they generally perform worse when modelling the quiescent population — particularly the number density of MQs (though note that there are significant uncertainties on observationally-inferred number densities; see e.g. \citealt{valentino20,valentino23}). For instance, \citet{lagos19} noted that the \citet{lagos18} (v1.0) version of \texttt{SHARK} matches the observed sub-mm number counts, but this version underpredicts the number density of MQs by $\sim1\,\rm{dex}$ relative to the updated \citet{lagos24} (v2.0) version, which better aligns with the lower limits of observational estimates. Similarly, \citet{chris21} showed that the original \texttt{Illustris} simulation reproduces sub-mm number counts while significantly underpredicting the quiescent population \citep{merlin19}. The same issue arises in \texttt{SIMBA}; while \citet{lovell21} demonstrated consistency with observed sub-mm number counts, \citet{merlin19} highlighted the underprediction of (in this case, intermediate mass) quiescent galaxy number densities across redshifts. The opposite trend is also found: \texttt{IllustrisTNG} and \texttt{EAGLE}, which better match the MQ population, systematically underpredict sub-mm number counts \citep[respectively]{chris21, cowley19}. These inconsistencies complicate our understanding of the formation and evolution of both population, particularly since DSFGs and MQs may be connected through evolutionary pathways \citep{daddi10, tacconi10, casey14, valentino20, chris21}. Resolving this tension is therefore a key challenge for theoretical astrophysics.

As mentioned above, one of the main reasons why modifying the IMF to solve the sub-mm number counts tension remains controversial is the high-dimensional parameter space of galaxy formation models. For example, SAMs typically have more than $10$ free parameters that are often manually tuned. This `calibration' process does not fully explore the range of possible scenarios and their physical implications, potentially obscuring alternative solutions. Consequently, robust calibration techniques are essential to rigorously test how well galaxy formation models reproduce observations. However, performing a comprehensive calibration is computationally expensive, as it requires extensive parameter-space exploration. In practice, this is infeasible for large-volume hydrodynamical simulations.

Unlike most models, the \texttt{L-Galaxies} SAM has incorporated a systematic calibration framework since \citet{henriques13}, using a Markov Chain Monte Carlo (MCMC) approach, known as the `MCMC mode'. This feature makes \texttt{L-Galaxies} uniquely flexible\footnote{The \citet{lagos24} version of \texttt{SHARK} also implemented a calibration method, but the final parameter choices were still refined via visual inspection.} by enabling calibration against multiple observables, including the stellar mass function, luminosity function, and quiescent fraction across different redshifts. In this work, we use the MCMC mode of the \citet{henriques20} version of \texttt{L-Galaxies} to systematically explore solutions to the SMG–MQ tension. Specifically, we calibrate the model using different sets of observational constraints, incorporating, for the first time, the number density of SMGs as a direct constraint. We then compare the galaxy properties predicted by the best-fit models, run on the \texttt{Millennium} simulation \citep{springel05}, across different calibration datasets. Finally, we analyze the dominant physical mechanisms driving these differences and assess the level of degeneracy in our most promising model.

The remainder of this paper is organized as follows. In Section \ref{sec:sam}, we provide an overview of the SAM used in this study. In Section \ref{sec:calib}, we describe our calibration framework, including the MCMC mode and the observational constraints used. We then present our model predictions, physical interpretations, and an analysis of degeneracies in Section \ref{sec:results}. In Section \ref{sec:discussion}, we discuss our results. We conclude with a summary of our findings in Section \ref{sec:summary}.

Throughout this work, we adopt the \citet{planck14} cosmology: $\sigma_8= 0.829$,
$H_0 = 67.3\,\rm{km\,s}^{-1}\,\rm{Mpc}^{- 1}$, $\Omega_{\Lambda}= 0.685$, $\Omega_m = 0.315$, $\Omega_b = 0.0487$, $f_b = 0.155$, and $n = 0.96$, consistent with the cosmologically rescaled version of the \texttt{Millennium} simulation \citep{angulo15}.

\section{Galaxy Formation Model} \label{sec:sam}

In this work, we use the \citet{henriques20} version of the \texttt{L-Galaxies} semi-analytic model (SAM) of galaxy formation. In this section, we briefly describe the principal aspects of this model.

The \texttt{L-Galaxies} SAM is optimized to run on the \texttt{Millennium} and \texttt{Millennium-II} N-body dark matter-only simulations \citep{springel05, boylan09}. In practice, \texttt{L-Galaxies} runs on the merger trees created with the \texttt{SUBFIND} algorithm \citep{springel01}. Additionally, \texttt{L-Galaxies} performs a cosmology scaling \citep{angulo10}, updating halo properties according to new cosmological parameters — in this case, the \citet{planck14} cosmology. After the cosmology scaling, the \texttt{Millennium} simulation volume is ($713.6\,\rm{cMpc})^{3}$ with a dark matter particle mass resolution of $m_p = 1.43  \times 10^9\,\rm{M_{\odot}}$. Here, we only run \texttt{L-Galaxies} on the \texttt{Millennium} simulation. 

The evolution of baryonic components is modelled by a set of differential equations that describe astrophysical processes. Initially, primordial gas begins to accrete onto sufficiently massive dark matter halos. The infalling gas is first added to the hot gas reservoir and subsequently transitions to the cold gas reservoir through radiative cooling. This version of \texttt{L-Galaxies} follows the evolution of cold gas in concentric rings within galaxies. This gas is further separated into HI and H$_2$, with only the latter forming stars, either through a secular process (based on H$_2$ surface density) or merger-induced starbursts. Besides triggering star formation, mergers are the main mechanism in the model for growing supermassive black holes (SMBHs), where SMBH mass growth is linked to the energy released from active galactic nuclei (AGN). AGN feedback is a crucial process regulating star formation in massive galaxies. Moreover, mergers also affect galaxy morphology, destroying disks and contributing to the growth or formation of the galaxy bulge.

Star formation and the evolution of the stellar component are related with various astrophysical processes in the galaxy evolution context. As stars reach their final stages, supernovae (SNe) and stellar winds release metals and energy into the interstellar medium (ISM) and circumgalactic medium (CGM). In this version of \texttt{L-Galaxies}, metal enrichment from AGB stars, SNe-Ia, and SNe-II is considered. Coupled with these events, the release of energy plays a crucial role in regulating subsequent star formation, i.e. SN feedback. The SN feedback in \texttt{L-Galaxies} operates in two ways: (re)heating and ejecting gas. The former (a) heats the cold gas within galaxies, transferring some to the surrounding CGM, and/or (b) reheats the CGM, thereby delaying cooling. On the other hand, when the energy release is significant, a fraction of the hot gas is ejected and later reincorporated after some time.

Environmental effects, such as tidal stripping, disruption, and ram pressure stripping, are also included in \texttt{L-Galaxies}. These processes occur when a halo is accreted by a more massive one. Among their effects on galaxies, these processes can remove hot gas atmospheres, modify galaxy components, and disrupt small systems.

The \citet{henriques20} version of \texttt{L-Galaxies} has 19 free parameters in total (Table 1 in \citealt{henriques20}), of which 15 were constrained using the MCMC mode of the model. Here, we follow the \citet{henriques20} configuration to calibrate the model, constraining the same $15$ free parameters.

To date, five more recent modifications of \texttt{L-Galaxies} have been published \citep{yates21, ayromlou21, izquierdo22, murphy22, spinoso23, yates24} since \citet{henriques20}. However, these versions introduce new treatments for specific astrophysical processes while still using the \citet{henriques20} version as a base. Therefore, in this work, we choose to use the default \citet{henriques20} version of \texttt{L-Galaxies}.

\section{Calibration Method} \label{sec:calib}
\subsection{The MCMC Mode} \label{sec:mcmc-mode}

First introduced by \citet{henriques09} and \citet{henriques10}, the Markov Chain Monte Carlo (MCMC) mode of \texttt{L-Galaxies} enables exploration of the model's free parameter space and its calibration against a set of observational data. Since the \citet{henriques13} version, the MCMC mode has operated on a representative sub-set of merger trees designed to approximate, as closely as possible, the predictions of observables — such as the stellar mass function, red/passive fractions, and number densities — when compared to those derived from the full cosmological volume. This approach significantly accelerates the MCMC process, making it feasible for calibrating galaxy formation models. However, not all halos in a given set of merger trees necessarily represent the overall predictions across all redshifts. Thus, a critical input for the MCMC mode is a carefully selected sub-sample of halos within the representative sub-set of merger trees at a given redshift, which, when combined, reproduce the results of the full model. Each halo in this sample is assigned a weight reflecting the number of similar halos in the entire simulation volume. In Section \ref{sec:sampling}, we introduce a new method for selecting both a sample of merger trees and the sub-sample of halos within those trees that effectively represent the predictions of the entire simulation.

The MCMC mode also requires the set of observables to constrain the model. The default \texttt{L-Galaxies} version already includes a large set of observables at different redshifts, both one-dimensional (such as SMFs, luminosity functions in different bands, and cold gas mass functions) and two-dimensional relations (such as black hole-bulge mass, stellar metallicity-stellar mass, and size-stellar mass relations). For instance, in \citet{henriques20}, the model was calibrated using the SMFs and fraction of quiescent galaxies both at $z = 0$ and $z = 2.0$, and the neutral hydrogen mass function at $z =0$ as observational constraints.
These observables are compared with the predictions generated at each MCMC step (from the sample of halos at a given redshift), and the likelihood is estimated. In Section \ref{sec:obsconst}, we describe the sets of observational constraints that we use in this work.\\
\indent In practice, multiple chains are run in parallel, each beginning from an initial point randomly displaced by a value $\sigma_{\rm initial}$. The Metropolis–Hastings algorithm is then applied using a log-normal Gaussian proposal distribution with width $\sigma$.

\subsection{Sample of Merger Trees} \label{sec:sampling}

As discussed in \citet{henriques13}, implementing an MCMC approach to calibrate galaxy formation models within cosmological simulations remains a computational challenge. In fact, the implementation of a robust calibration is still not feasible for hydrodynamical simulations. This is primarily due to the need to track the evolution of millions of galaxies while repeatedly varying the model's free parameters. A key alternative, introduced in \citet{henriques09} and \citet{henriques10}, involves using sub-volumes of the simulation. While this approach improves efficiency, the sub-volumes may not fully represent the predictions of the entire simulation. Specifically, probes of rare populations, such as the massive end of the stellar mass function and the fraction of quiescent galaxies, can be under- or over-predicted in certain regions.

To address this, \citet{henriques13} developed a method for selecting samples of merger trees that accurately represent the predicted luminosity function of a fiducial model. This approach not only improves the representativeness of the sample but also significantly reduces computational time, as far fewer halos and merger trees are needed to reproduce the predictions of the full volume. However, initial tests performed as part of this project showed that their method fails to adequately represent the number density of SMGs — a key observable for this study — primarily due to the rarity of these galaxies. This limitation might also apply to other non-conventional observables that probe rare populations. Therefore, the selection of merger tree samples should be based on the specific observational constraints required for the study.

\begin{figure}
    \centering
    \includegraphics[width=\columnwidth]{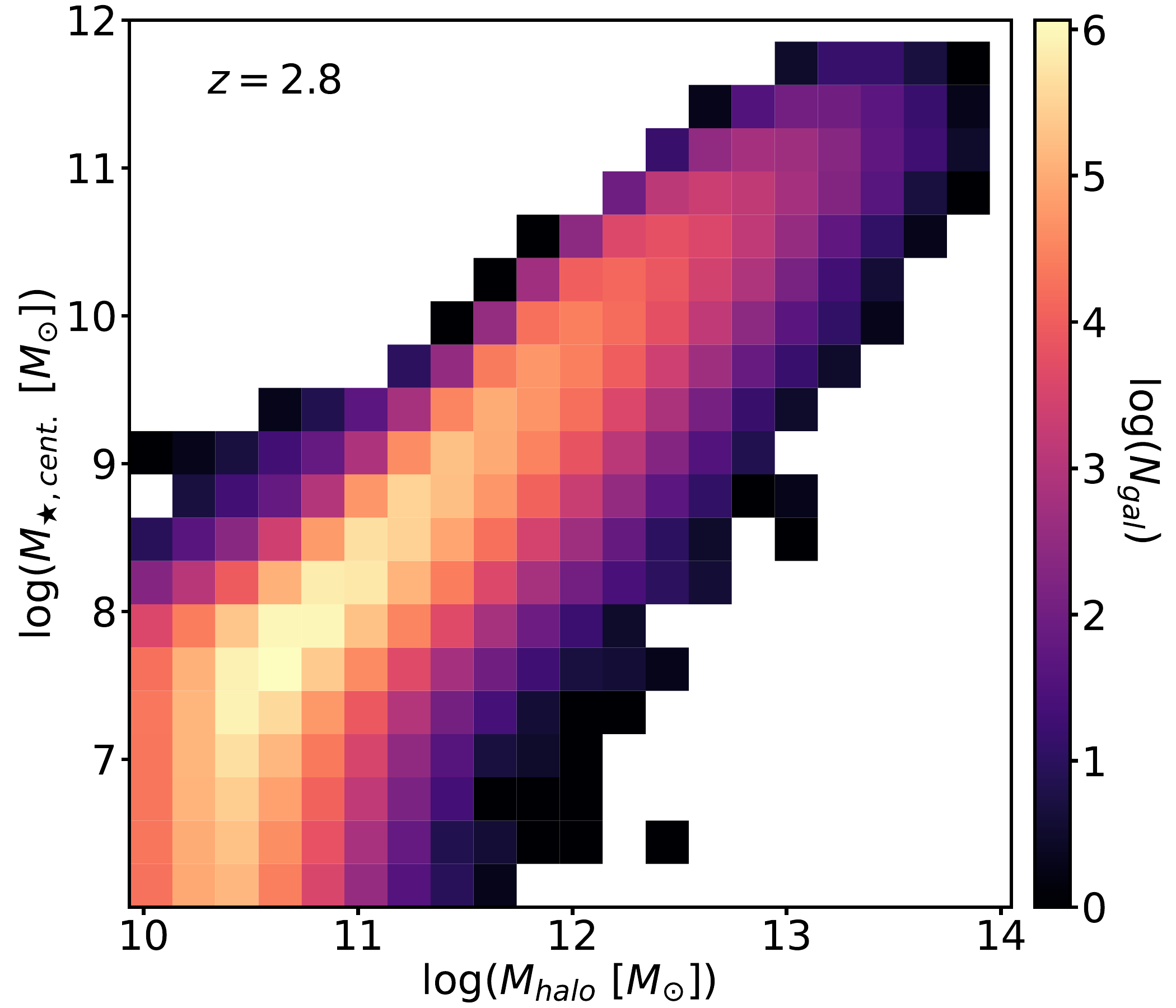}
    \caption{The stellar mass - halo mass relation for central galaxies in a $20\times20$ grid. We use this distribution to obtain a preliminary sample of dark matter halos and identify an optimal sample merger trees.}
    \label{fig:2dgrid}
\end{figure}

Here, we introduce a new method for selecting samples of merger trees that can also be applied to generate representative samples for other observables used in the calibration of galaxy formation models. This method involves two main steps. The first step involves selecting a reasonable sub-sample of merger trees consistent with the observable predictions of the entire simulation at a given redshift. We choose $z=2.8$ in this work, as our motivation is to test whether we can match the SMG density and the quiescent galaxy fraction at this epoch. In particular, our observables, which we describe in Section \ref{sec:obsconst}, are the stellar mass function (SMF), the quiescent galaxy fraction as a function of stellar mass ($f_{\rm Q}$), the SMG number density ($n_{\rm SMG}$), and the neutral hydrogen mass function (HIMF). To achieve this, starting with a preliminary (fiducial) model that predicts a higher number of SMGs than the default 2020 version of \texttt{L-Galaxies}, we construct a 2-D grid ($20\times20$ bins) of the virial mass–stellar mass relation for all central galaxies (galaxies containing the most-bound dark matter particle in each FOF halo group) at $z=2.8$, as shown in Figure \ref{fig:2dgrid}.

We could also include other key galaxy properties in this selection process, such as star formation rate, local overdensity, or metallicity. However, increasing the number of dimensions also increases the number of cells required to select a representative sample of halos. For this reason, we chose to consider only the halo and stellar masses of the central galaxy, which is sufficient to obtain a representative sample, as shown below.
Notice that we set a $M_{\star} = 10^6\,{\rm{M_{\odot}}}/h$ lower limit in Figure \ref{fig:2dgrid}, which is below the stellar mass of well-modelled galaxies due to mass resolution, when run on \texttt{Millennium}. This ensures the sample of halos we select represents the predictions of the entire simulation volume. The upper limits are set as the maximum stellar and halo masses.

We start our procedure by randomly selecting two halos in each non-zero cell of the 2-D grid. This grid has $215$ non-zero cells. Hence, we select $430$ FOF halos/central galaxies. Then, we assign the weight of each halo as the number of selected halos divided by the total number in its respective cell. Our algorithm \textit{accepts} a sample of halos if the sum of the average relative errors between the sampled observables and the full predictions is lower than the set of halos previously accepted. After testing 20,000 sets of halos, the algorithm retrieves the set with the lowest sum of average relative errors. This first step results in a representative sample of merger trees.

As mentioned earlier, it is not necessarily the case that the observables from all halos in the set of merger trees are consistent with the entire volume predictions at all redshifts. Therefore, the second step in this process involves selecting a sub-sample of halos (from the selected merger trees) at all redshifts separately. Note that \texttt{L-Galaxies} runs over the sample of merger trees, and the runtime is almost independent of the halo sub-sample size.
In this step, halos are selected based only on their virial mass. To find the optimal number of mass bins, we choose the highest number that produces histograms without empty bins (from the set of merger trees) within the mass limits. The procedure starts by testing only one halo per mass bin. If no better sample is found after $50$ trials, the number of selected halos per bin increases\footnote{If the mass bin contains fewer halos than requested, we use all of them without repetition.}. This algorithm ends after $20,000$ tests. Note that the number density of SMGs at $z \lesssim 1.5$ is very low, so we do not include this observable when estimating representativeness for $0 < z < 1.5$.

In general, the average relative errors are $\sim 10\%$, $\sim 25\%$, $\sim 5\%$, and $\sim 20\%$ for SMFs, $f_{\rm Q}$s, $n_{\rm SMG}$, and HIMF, respectively. We compare the sampled and full-volume predictions in Appendix \ref{sec:sampling_comp}. This procedure yielded a sample of merger trees/halos representing all the observables we will use to constrain our model, without significantly increasing runtime.

\subsection{Observational Constraints} \label{sec:obsconst}

A crucial input to the \texttt{L-Galaxies} MCMC mode is the set of observational constraints, the observables, as the algorithm compares the proposed model at each MCMC step to this dataset. Here, we describe the updates to the observational data used in this study.

\subsubsection{Stellar Mass Function} \label{sec:obs_smfs}
\citet{henriques15, henriques20} calibrated the free parameters of \texttt{L-Galaxies} using combined stellar mass functions from SDSS at $z=0$ \citep{baldry08,baldry12,li09} and ULTRAVISTA at $z =2$ \citep{ilbert13,muzzin13}. In contrast, we use the stellar mass functions derived by \citet{leja20} and the quiescent fractions presented by \citet{leja22}. The primary motivation for this change is that the stellar mass functions of \citet{leja20} provided a resolution to the tension between the observed star formation rate density and the stellar mass density \citep{madau14, leja15, tomczak16}, which is evident in many datasets commonly used for calibration \citep[e.g.][]{baldry12,ilbert13,muzzin13, tomczak14}. We provide a summary of these datasets here. 

\citet{leja20} employed the \texttt{Prospector} SED fitting code \citep{leja17, leja19b}, which uses non-parametric star formation histories (SFHs) to construct modelled SEDs. As shown by \citet{leja19a}, non-parametric SFHs recover input SFHs with significantly less bias compared to parametrized SFHs. \citet{leja19b} demonstrated that this method yields stellar masses approximately $0.1-0.3\,\rm{dex}$ larger and total SFRs approximately $0.1-1\,\rm{dex}$ lower than previous studies, suggesting a reconciliation between these two observables. \citet{leja20} developed a model to describe the evolution of the SMF (referred to as the ``continuity model") by fitting SMFs derived from the 3D-HST \citep{skelton14} and COSMOS2015 \citep{laigle16} surveys, in the range $0.2 < z < 3.0$. Given the redshift limits of the model, we use the continuity model to derive the SMF at $z = 0.4$ and $z = 2.8$. To achieve this, we followed the procedure outlined in Appendix B of \citet{leja20} to generate the posterior distribution of the median SMF at each redshift and its associated $1\,\sigma$ uncertainty.

\begin{figure} 
    \centering 
    \includegraphics[width=\columnwidth]{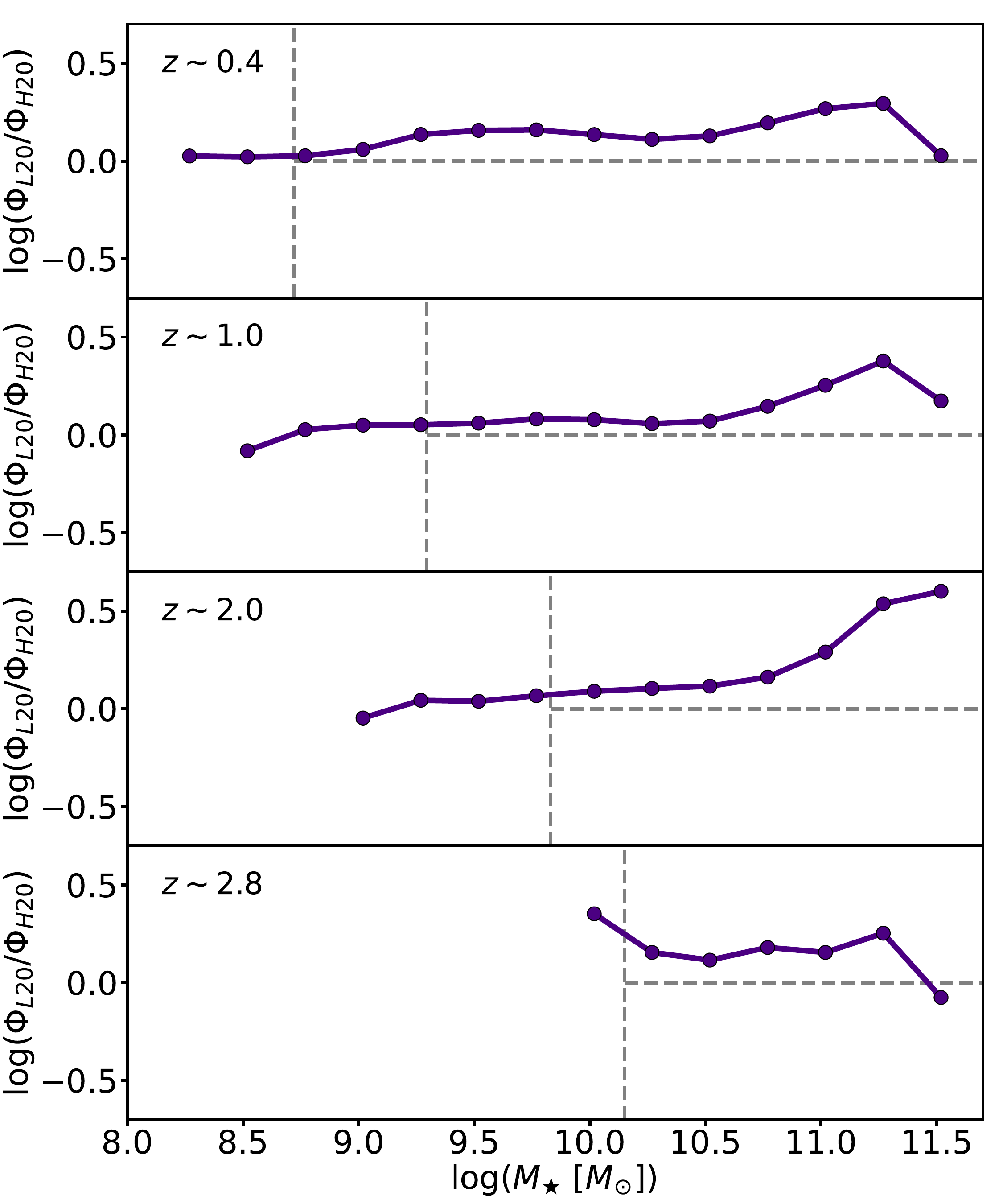} 
    \caption{Comparison between the stellar mass functions of \citet{leja20} (continuity model used in this work; L20) and those used in \citet{henriques20} (compilation from literature; H20) at $z \sim$ 0.4, 1.0, 2.0, and 2.8. Dashed grey horizontal and vertical lines indicate unity and the mass completeness of the \citet{leja20} dataset, respectively. The L20 stellar mass functions indicate higher galaxy number densities, particularly at the massive end.} 
    \label{fig:smf_comp} 
\end{figure}

Figure \ref{fig:smf_comp} illustrates the differences between the SMFs used in this work and those used by \citet{henriques15} and \citet{henriques20}. Across almost the entire stellar mass range above the completeness limit, the SMFs derived from the \citet{leja20} continuity model (L20, in Figure \ref{fig:smf_comp}) exceed the dataset used by \citet{henriques15} and \citet{henriques20} (H20) by $0.1 - 0.2\,\rm{dex}$. This difference generally increases with stellar mass, except at the most massive bin, where the number densities are comparable.

\subsubsection{Quiescent Fraction}
Following \citet{henriques15} and \citet{henriques20}, we used the fraction of `red/quiescent' galaxies as a function of stellar mass at different redshifts to calibrate the model. In both of these earlier studies, data from \citet{muzzin13}, \citet{ilbert13}, and \citet{tomczak14} were combined with $UVJ$ colour-colour criteria applied to define `red' galaxies. At $z = 0$, the \citet{baldry04} $u - r$ colour cut was used. However, as discussed in \citet{rodriguez17}, systematic effects — such as the assumed initial mass function (IMF), stellar population synthesis (SPS) model, photometric calibrations, and dust extinction — can bias the models when comparing different datasets. For simplicity and to maintain a homogeneous sample, we used the \citet{leja22} data to construct this observational constraint. This dataset is the same as that used in \citet{leja20} to derive the continuity model of the SMF.

Additionally, we replaced the fraction of `red' (using $UVJ$ colour-colour criteria) galaxies with the fraction of `quiescent' galaxies, defining them using a specific star formation rate (sSFR) threshold of $\log (\text{sSFR}/\text{yr}^{-1}) \leq -11$. This choice was motivated by the need for a consistent selection at different redshifts, as well as the need to avoid systematic biases introduced by the different SPS models used to compute galaxy magnitudes. For instance, \citet{leja20} and \citet{leja22} used the Flexible Stellar Population Synthesis (\texttt{FSPS}) code \citep{conroy09}, whereas the available SPS models used in \texttt{L-Galaxies} include \citet{bruzual03}, \citet{maraston05}, and \citet{charlot07}. Instead, we quantified the fraction of galaxies with $\log (\text{sSFR}/\text{yr}^{-1}) \leq -11$ in five stellar mass bins ($\log (M_{\star} / \rm{M_{\odot}}) = [9.0-9.5, 9.5-10.0, 10.0-10.5, 10.5-11.0, 11.0-12.0]$) at $z = 0.4$, $1.0$, $2.0$, and $2.8$. However, we calibrated the model only using the quiescent fraction ($f_{\rm Q}$) at two redshifts, $z = 0.4$ and $2.8$.

In order to obtain robust estimates, for each galaxy in the \citet{leja22} dataset, we sampled $10,000$ values based on the uncertainties in (1) redshift, (2) stellar mass, and (3) sSFR, assuming a Gaussian distribution centered on the most likely value and with $\sigma$ equal to the uncertainties in these properties. We then estimated the median of the $10,000$ sampled $f_{\rm Q}$ values and the associated $1\,\sigma$ error. $f_{\rm Q}$ is also influenced by sample size, so we computed the total error in $f_{\rm Q}$ as the quadratic sum of the uncertainty-induced error and the Poissonian error.

\subsubsection{Number Densities of sub-mm galaxies}
As shown in \citet{yo24}, the \citet{henriques15} version of \texttt{L-Galaxies} underpredicts the sub-mm number counts when using the \citet{rachel23} scaling relations to model the observed-frame $870\,\mu$m flux densities ($S_{870}$; since \texttt{L-Galaxies} does not make predictions for sub-mm fluxes, $S_{870}$ was modelled in that work as a function of SFR, $M_{\star}$, $M_{\rm dust}$, and redshift, based on detailed radiative transfer post-processing on highly-resolved zoom-in galaxies; see \citealt{rachel23}). This underprediction of sub-mm number counts persists in the \citet{henriques20} version of the SAM. Motivated by obtaining a better match to observationally-derived sub-mm number counts, we included SMG number density measurements as an additional observational constraint.

Ideally, the full sub-mm number counts would be used as observational constraints. However, the MCMC mode estimates likelihoods by comparing the proposed model to the observables at a few specific snapshots (redshifts). For simplicity, we instead use the SMG number density as an observational constraint. We estimated the SMG number density using the \citet{dud20} catalog, which provides photometric redshifts for SMGs observed in the AS2UDS survey \citep{simpson17,stach18,stach19,dud20}. This catalogue consists of $870\,\mu\rm{m}$ continuum ALMA follow-up observations of SCUBA-2 detections in the UDS field (S2UDS; \citealt{geach17}), covering an area of $0.96\,\rm{deg}^2$. The ALMA survey targeted S2UDS sources with $4\,\sigma$ detections (i.e. $S_{850}\geq 3.6\,\rm{mJy}$), but the sample is incomplete at these flux densities (see \citealt{geach17}). Therefore, we adopted $S_{870} = 5.2\,\rm{mJy}$ (where completeness exceeds $90$ per cent) as the flux density threshold for our calibration. 

To estimate the number density, we selected all SMGs with photometric redshifts within $z_c \pm \Delta z$, where $z_c = 2.8$ (our highest redshift for SMF and $f_{\rm Q}$), and $\Delta z = 0.35$. Similar to our approach for $f_{\rm Q}$, we sampled $10,000$ values of redshift and $S_{870}$ flux density for each galaxy, accounting for the uncertainties in these estimates. The SMG number density was calculated as the median of the $10,000$ sampled values, with the observational uncertainty taken as the standard deviation. Again, we included the Poissonian error contribution due to the sample size. Finally, the number density of galaxies with $S_{870} \geq 5.2\,\rm{mJy}$ at $z \sim 2.8$ is $n_{\rm SMG} = (2.48 \pm 0.3) \times 10^{-5}\,h^{3}\,\text{Mpc}^{-3}$.

To implement this new observable as a constraint, we modified the \texttt{L-Galaxies} code to estimate $S_{870}$ using the \citet{rachel23} scaling relations and, then, estimate the SMG number density. Since the \citet{rachel23} scaling relations are parametrized by the average star formation rate over the last $10\,\rm{Myr}$, we used the instantaneous SFR (\texttt{SfrInst}) output from \texttt{L-Galaxies}. However, this SFR does not account for the contribution from merger-induced starbursts. Therefore, we further modified \texttt{L-Galaxies} to include this factor in the \texttt{SfrInst} parameter.

\subsubsection{Neutral Hydrogen Mass Function}

Like \citet{henriques20}, we included the HI Mass Function at $z = 0$ as an additional constraint, alongside the SMF and $f_{\rm Q}$ at two higher redshifts. The observational data combine results from \citet{zwaan05}, \citet{haynes11}, and \citet{jones18}. Recall that in this study, we use SMFs and $f_{\rm Q}$s at $z = 0.4$ and $z = 2.8$, rather than at $z = 0$ and $z = 2.0$ as in \citet{henriques20}, as explained earlier in this section.

\begin{table*}
\caption{List of MCMC configurations used throughout this paper.}
\centering
\begin{tabular}{ccccccc}
\hline
Config & \multicolumn{2}{c}{SMF} & \multicolumn{2}{c}{$f_{\rm Q}$} &  $n_{\rm SMG}$ & HIMF \\ 
 & $z=0.4$ & $z=2.8$ & $z=0.4$ & $z=$2.8 & $z=2.8$ & $z=0$ \\
\hline
0: base$_{\rm H20}$ & \checkmark & \checkmark & \checkmark & \checkmark & &\checkmark \\
1: base$_{\rm L20}$ & \checkmark & \checkmark & \checkmark & \checkmark & &\checkmark \\
2: all(base$_{\rm L20}$+$n_{\rm SMG}$) & \checkmark & \checkmark & \checkmark & \checkmark & \checkmark &\checkmark \\ 
3: no hi-$z$ SMF& \checkmark &  & \checkmark & \checkmark & \checkmark &\checkmark \\
4: no hi-$z$ $f_{\rm Q}$ & \checkmark &  \checkmark & \checkmark &  & \checkmark &\checkmark \\
5: no hi-$z$ SMF, $f_{\rm Q}$ & \checkmark &   & \checkmark &  & \checkmark &\checkmark \\
6: no HI MF & \checkmark &  \checkmark & \checkmark & \checkmark & \checkmark &  \\
7: no low-$z$ $f_{\rm Q}$ & \checkmark & \checkmark & & \checkmark & \checkmark & \checkmark \\
8: no low-$z$ SMF, $f_{\rm Q}$ &  & \checkmark & & \checkmark & \checkmark & \checkmark \\
\hline
\end{tabular}
\label{tab:configs}
\end{table*}

\subsection{MCMC Configurations}

In order to assess how sensitive \texttt{L-Galaxies} is to the observational constraints used for calibration, we ran the \texttt{L-Galaxies} MCMC mode for nine different sets of constraints. These sets are referred to as configurations throughout this work. In principle, we expect to obtain a different model for each configuration. Understanding which physical models favor specific observables is crucial to identifying the key discrepancies between galaxy formation models and observations.

The configurations tested in this work are listed in Table \ref{tab:configs}. The observables (SMFs, $f_{\rm Q}$, HIMF, and $n_{\rm SMG}$) at a given redshift ($z =0.4$ and $2.8$) used as constraints for the different configurations are denoted by the check marks. Note that for configuration base$_{\rm H20}$, we used the same constraints as in \citet{henriques20}.

\subsection{Running the MCMC mode} \label{sec:run}

We ran the \texttt{L-Galaxies} MCMC mode with the sample of merger trees obtained as described in Section \ref{sec:sampling}, using the observational data detailed in Section \ref{sec:obsconst}, and for the nine configurations listed in Table \ref{tab:configs}. The free parameters of the model were initially randomly sampled with a standardized space displacement from the starting point of $\sigma_{\rm initial} = 0.1$, and thereafter randomly sampled from a log-normal distribution with $\sigma = 0.25$ as in \citet{henriques20}. We used a modified version of the MCMC mode that compares the likelihood at each step in logarithmic space, improving the efficiency of convergence. This is necessary to avoid numerical underflow, which can round likelihoods to zero due to the high dimensionality of the likelihood space and the presence of multiple observational constraints. In some cases, we encountered likelihood values on the order of $\gtrsim 10^{-100}$.

For each configuration, we ran the MCMC mode with $96$ chains for approximately $5000$ steps. Although the number of steps is lower compared to traditional MCMC fitting, the high number of chains ensures that the free parameter space is thoroughly explored. Tests confirmed convergence: we did not find any new accepted point (with a higher likelihood) within the final $\gtrsim$1000 steps. After obtaining an initial best-fit model (i.e. the set of parameters with the highest likelihood) for each configuration, we performed an additional run of $2000$ steps. For these new runs, we set the starting point to the previously obtained best-fit parameters and sampled the proposed parameters with $\sigma_{\rm initial} = 0.05$ and $\sigma = 0.15$.

For each configuration, the MCMC process consumed approximately $110,000$ CPU hours. In total, this work used around $1$ million CPU hours. The $15$ best-fit parameters obtained for each configuration are presented in Appendix \ref{sec:best-fits}.

\section{Results} \label{sec:results}
In this section, we present the predictions of the galaxy properties obtained by the best fit of each configuration (Section \ref{sec:props}), alongside key figures that aid in interpreting the main physical aspects of the models (Section \ref{sec:physcs}). 

\subsection{Predictions for Galaxy Properties} \label{sec:props}
After obtaining a best-fit model from each configuration, we ran \texttt{L-Galaxies} for the entire \texttt{Millennium} volume following the procedure described in Section \ref{sec:run}. Here, we present the main predictions, starting with the properties used to calibrate the model.

\begin{figure*}
    \centering
    \includegraphics[width=\textwidth]{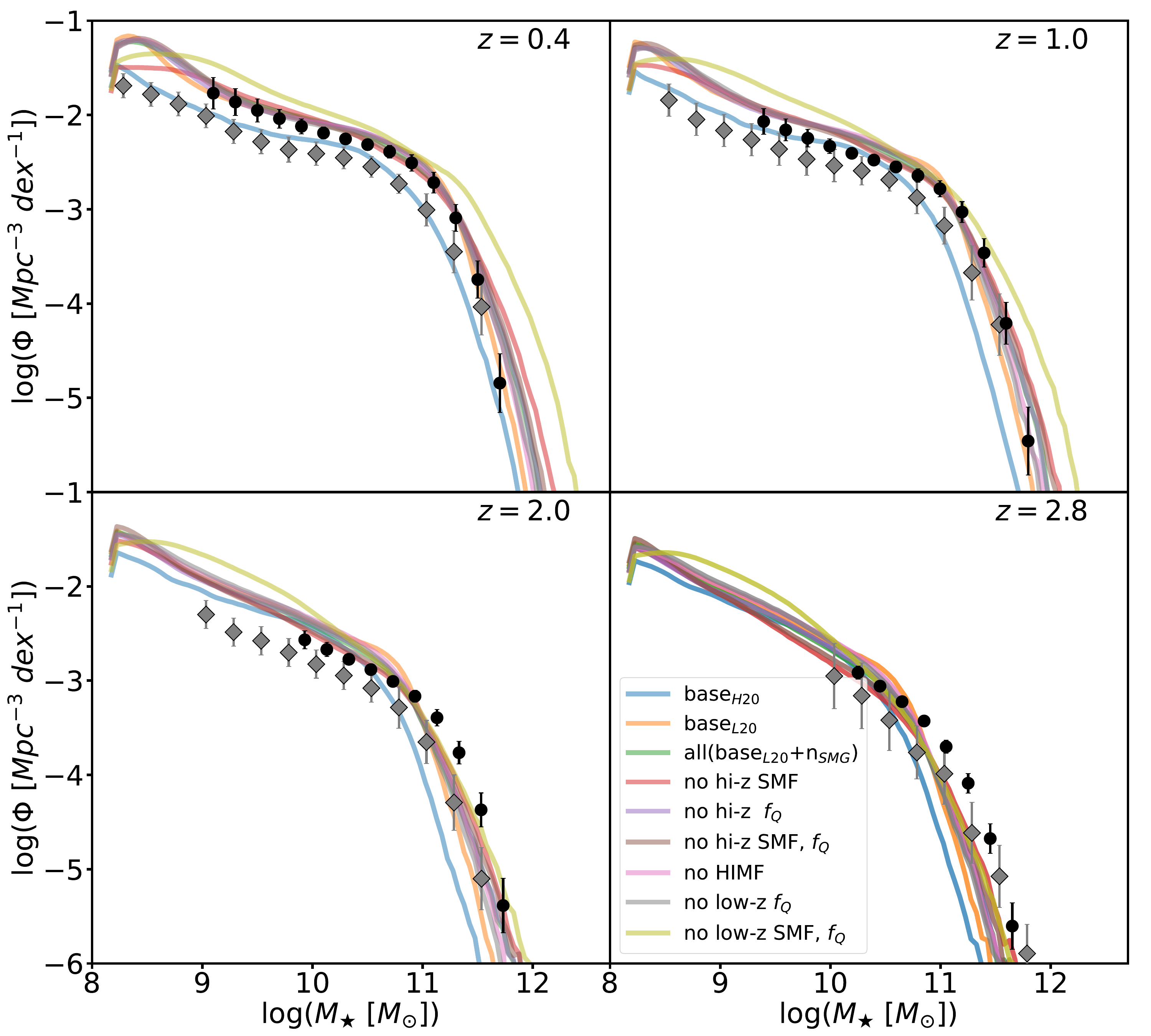}
    \caption{The predicted stellar mass functions (SMFs) at $z = 0.4$, $1.0$, $2.0$, and $2.8$, for every configuration listed in Table \ref{tab:configs}, compared to the \citet{leja20} continuity model (black dots) and the dataset used in \citet{henriques20} (grey diamonds). Overall, the configurations predict consistent SMFs with the adopted observational constraints at $z = 0.4$ and $1.0$. However, they slightly underpredict the observed massive end at $z = 2.0$ and $2.8$.}
    \label{fig:smfs}
\end{figure*}

\subsubsection{Predicted Stellar Mass Functions}
In Figure \ref{fig:smfs}, we show the stellar mass functions for each calibrated configuration at $z = 0.4$, $1.0$, $2.0$, and $2.8$. SMF data at $z = 0.4$ and $2.8$ were used to calibrate some configurations. The figure includes two observed SMFs: the SMF used to calibrate the \citet{henriques20} model (grey diamonds) and the SMF derived by \citet{leja20} (black dots; see Section \ref{sec:obs_smfs} for details). As expected, the predicted SMFs for the `base$_{\rm H20}$' configuration are lower across all redshifts compared to other configurations, as the \citet{henriques20} dataset was used for calibration.

Overall, our predictions match the observationally-inferred SMFs well at $z = 0.4$ and 1.0 but underpredict the number densities at $z = 2.0$ and 2.8, particularly at the massive end. Another noteworthy result is the prediction from the `no low-$z$ SMF, $f_{\rm Q}$' configuration, which significantly overestimates the number density of massive galaxies at low redshift, due to the lack of constraints there.

\begin{figure*}
    \centering
    \includegraphics[width=\textwidth]{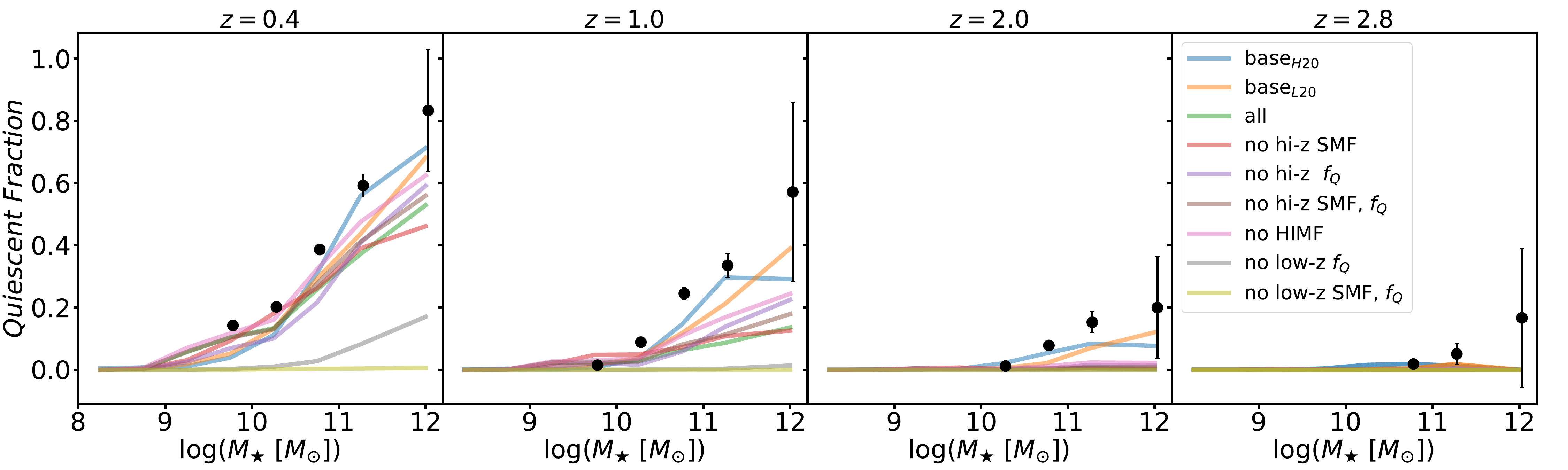}
    \caption{The predicted quiescent fraction ($f_{\rm Q}$; $\log ({\rm sSFR}/{\rm yr}^{-1}) \leq$ -11) as a function of stellar mass at $z = $ 0.4, 1.0, 2.0, and 2.8, for every configuration listed in Table \ref{tab:configs}, compared to \citet{leja22} data (black dots). Configurations where the number density of SMGs is not input as an observational constraint (`base$_{\rm H20}$' and `base$_{\rm L20}$') are in better agreement with the observational data. When $f_{\rm Q}$ at low-$z$ is not a constraint (`no low-$z$ $f_{\rm Q}$' and `no low-$z$ SMF; $f_{\rm Q}$'), the model critically underpredicts this population. }
    \label{fig:fqs}
\end{figure*}

\subsubsection{Predicted Fraction of Quiescent Galaxies}

The second main observational constraint used in this work is the fraction of quiescent galaxies in stellar mass bins, $f_{\rm Q}$, defined by $\log(
\rm{sSFR/yr}^{-1})<-11$  for all configurations except `base$_{\rm H20}$'\footnote{For this configuration, the default \citet{henriques20} definition was retained exclusively for the calibration process.}. Figure \ref{fig:fqs} shows the $f_{\rm Q}$ predictions at $z = 0.4$, $1.0$, $2.0$, and $2.8$.

Despite using a different definition of quiescent galaxies during the calibration of the `base$_{\rm H20}$' configuration, its $f_{\rm Q}$ predictions are similar to its counterpart (`base$_{\rm L20}$'), which was also calibrated with different observational data. Both configurations (`base$_{\rm H20}$' and `base$_{\rm L20}$') provide the best match to observational results up to $z = 2.0$. As expected, when $f_{\rm Q}$ at low redshift is not used as a constraint (in configurations `no low-$z$ $f_{\rm Q}$' and `no low-$z$ SMF, $f_{\rm Q}$'), the fraction of quiescent galaxies is significantly underpredicted across all redshifts analyzed. This highlights the importance of including $f_{\rm Q}$ as a calibration constraint. In general, the other configurations, which incorporated the number density of SMGs as a constraint, tend to underpredict $f_{\rm Q}$, with the discrepancy being most significant at $z = 2.0$ and 2.8. Among these, the `no HIMF' configuration performs the best in predicting $f_{\rm Q}$.

\begin{figure}
    \centering
    \includegraphics[width=\columnwidth]{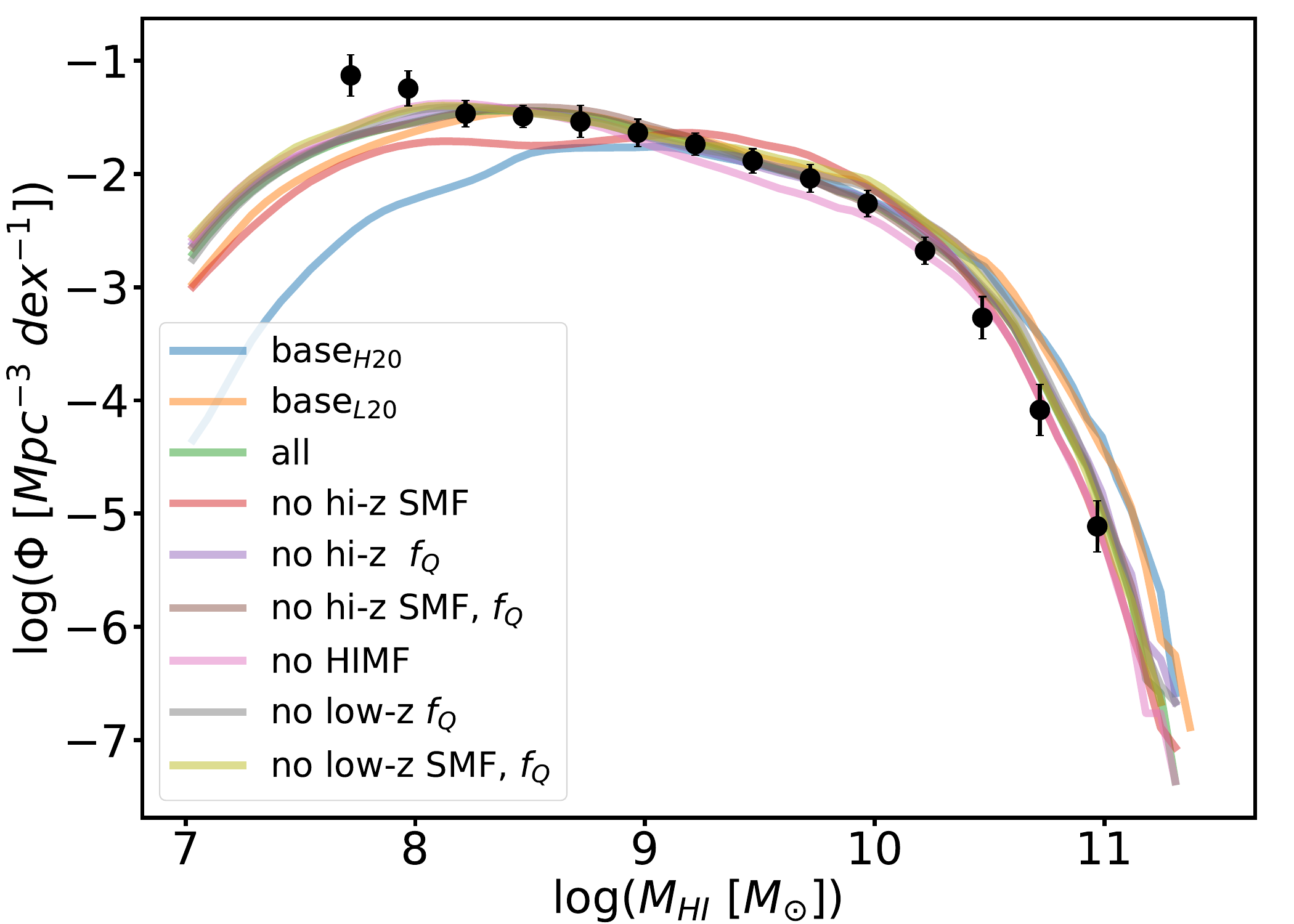}
    \caption{ The predicted neutral hydrogen mass function (HIMF) at $z = $ 0 for every configuration listed in Table \ref{tab:configs}, compared to the observational data used as a constraint (black dots; compilation of results from \citealt{zwaan05}, \citealt{haynes11}, and \citealt{jones18}). Overall, all predicted HIMFs are similar and consistent with the observational data. The main differences are at the low-mass and high-mass end of the distributions. Configurations where $n_{\rm SMG}$ is not used as a constraint (`base$_{\rm H20}$' and `base$_{\rm L20}$') show an excess of galaxies with large HI reservoirs. Although the HIMF was not a constraint for configuration `no HIMF', its prediction agrees with the observed HIMF.}
    \label{fig:HImf}
\end{figure}

\subsubsection{Predicted HI Mass Function}

Following \citet{henriques20}, we use the neutral hydrogen mass function (HIMF) at $z = 0$ as an observational constraint in all configurations except `no HIMF'. Figure \ref{fig:HImf} shows our predictions. Overall, the best-fit models successfully reproduce the observed HIMF and exhibit similar distributions. However, notable differences arise at both the low-mass and high-mass ends of the distribution. For instance, the `base$_{\rm H20}$' configuration significantly underpredicts the number density at the low-mass end compared to other configurations and, like `base$_{\rm L20}$', shows an excess at the high-mass end. The strong downturn at low HI mass may be driven by the resolution limit of \texttt{Millennium}, which is $\sim 10^{9.5}\,\rm{M_{\odot}}$.

Interestingly, despite the `no HIMF' configuration not being constrained by the HIMF, its predictions match the observational data well. This result suggests that the HIMF may not be a critical observational constraint.

\subsubsection{Predicted SMG population}
\begin{figure}
    \centering
    \includegraphics[width=\columnwidth]{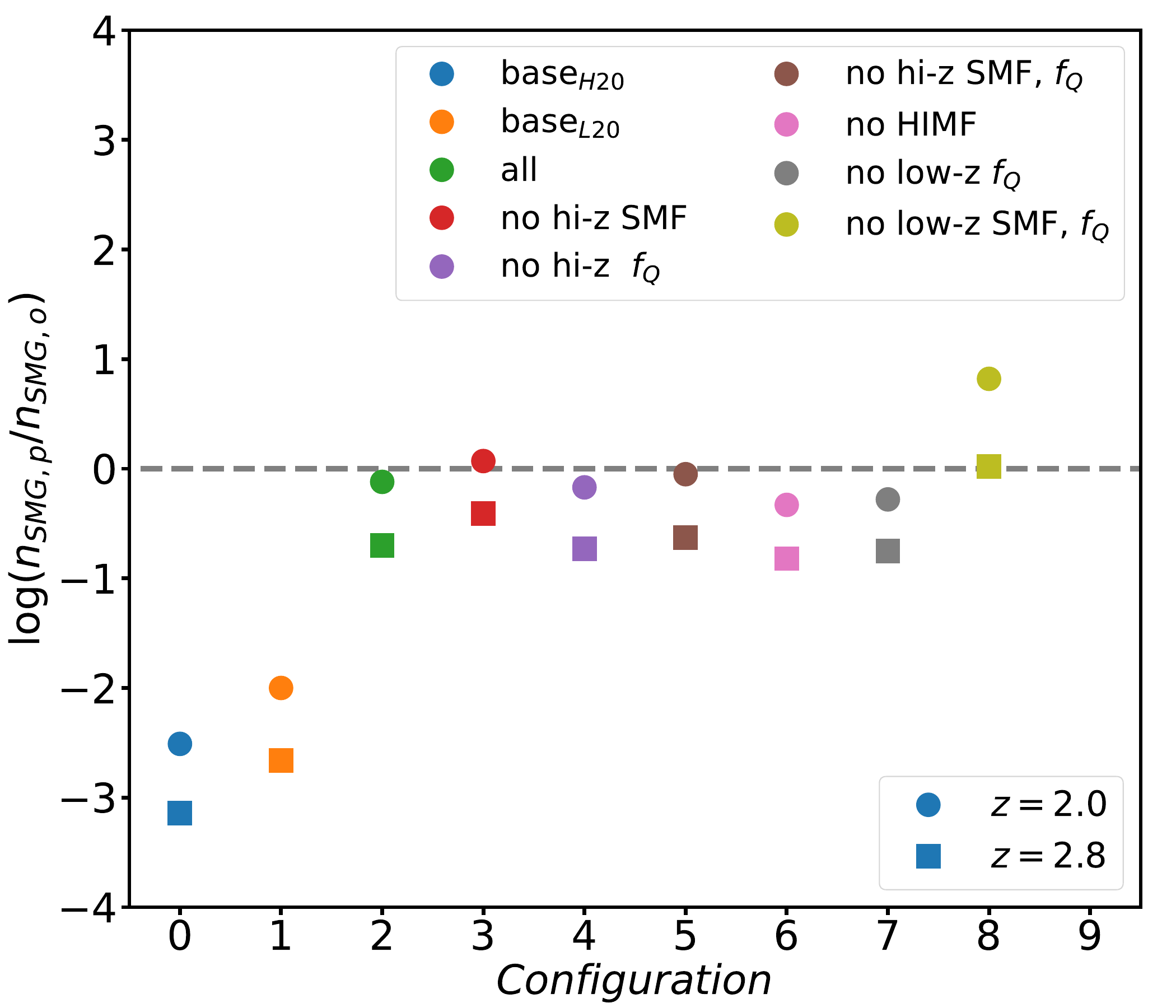}
    \caption{Deviation of the predicted number density of SMGs, $n_{\rm SMG,p}$, from the observationally-inferred number density, $n_{\rm SMG,o}$ at $z= 2.0$ (dots) and $z=2.8$ (squares), for our various models listed in Table \ref{tab:configs}. Our new models that include $n_{\rm{SMG}}$ at $z=2.8$ as a constraint match observational estimates significantly better than those that do not (orange and blue symbols). The observed SMG number densities are best-matched by the `all', `no hi-$z$ SMF' and `no hi-$z$ SMF, $f_{\rm{Q}}$' configurations. }
    \label{fig:nsmgs}
\end{figure}
%\begin{table}
%\caption{Deviation of the predicted number density of SMGs, $n_{\rm SMG,p}$, from the observationally-inferred number density, $n_{\rm SMG,o}$ at $z=$ 2.0, and 2.8, for our various models. The observed SMG number densities are best-matched by the `all', `no hi-z SMF' and `no hi-z SMF, $f_{\rm{Q}}$' configurations. }
%\centering
%\begin{tabular}{ccc}
%\hline 
%Config & \multicolumn{2}{c}{$\log(n_{\rm SMG,p}/n_{\rm SMG,o} )$} \\ 
% & $z=$2.0 & $z=$2.8 \\
%\hline
%base$_{H20}$ & -2.51 & -3.14 \\
%base$_{L20}$ & -2.00 & -2.66 \\
%all & -0.12 & -0.70 \\
%no hi-z SMF & 0.07 & -0.41 \\
%no hi-z  $f_{Q}$ & -0.17 & -0.73 \\
%no hi-z SMF, $f_{Q}$ & -0.05 & -0.63 \\
%no HIMF & -0.33 & -0.82 \\
%no low-z $f_{Q}$ & -0.28 & -0.75 \\
%no low-z SMF, $f_{Q}$ & 0.82 & 0.02 \\
%\hline
%\end{tabular}
%\label{tab:nsmgs}
%\end{table}

The primary novelty of this work is the inclusion of the number density of bright sub-millimeter galaxies, defined here as all galaxies with $S_{870} \geq$ 5.2 mJy, as an observational constraint in the calibration process. Our main motivation for adding this constraint is to better match the observed sub-mm number counts. We first present our predictions for the SMG number densities, $n_{\rm SMG}$, at $z = 2.0$ and $z = 2.8$ compared to observational data in Figure \ref{fig:nsmgs}. It is important to note that only the $z = 2.8$ $n_{\rm SMG}$ was used to constrain all models, except for the `base$_{\rm H20}$' and `base$_{\rm L20}$' configurations, where no SMG constraints are used in the calibration. Both of these configurations critically underpredict the SMG number densities by more than two orders of magnitude, as has been seen in many previous works (see Section \ref{sec:intro}). This underscores the importance of this constraint for simultaneously reproducing the SMG and quiescent populations.\\
\indent Despite the inclusion of this constraint, all configurations except `no low-$z$ SMF $f_{\rm Q}$' underpredict the $n_{\rm SMG}$ at $z = 2.8$ by at least a factor of 2.5 (e.g. `no hi-$z$ SMF'). This discrepancy could be explained by the underprediction of the massive end of the stellar mass function (SMF) at $z = 2.8$. In contrast, our predictions for $z = 2.0$ align more closely with observational data. The worst-performing configuration, `no HIMF', underpredicts the SMG number density at $z = 2.0$ by a factor of $\sim 2.1$.\\
\indent The configurations that best match the observed SMG number densities are `no hi-$z$ SMF', `no hi-$z$ SMF, $f_{\rm Q}$', and `all'. The `no hi-$z$ SMF' configuration slightly overpredicts the number density at $z = 2.0$ and is the closest match at $z = 2.8$, while the `no hi-$z$ SMF, $f_{\rm Q}$' configuration underpredicts the $z = 2.0$ number density by only a factor of $1.12$. In general, the configuration `all' presents similar number densities to configuration `no hi-$z$ SMF, $f_{\rm Q}$' but is slightly lower at both redshifts. Interestingly, the configuration `no low-$z$ SMF, $f_{\rm Q}$' is the only one that matches the observed $n_{\rm SMG}$ at $z = 2.8$, but it overpredicts the $z = 2.0$ $n_{\rm SMG}$ by a factor of $6.6$. Notably, this configuration does not exhibit significant differences in the SMFs or $f_{\rm Q}$ at high redshift ($z = 2.0$ and $z = 2.8$) compared to other configurations, nor does it differ in the HIMF predictions at $z = 0$.

\begin{figure}
    \centering
    \includegraphics[width=\columnwidth]{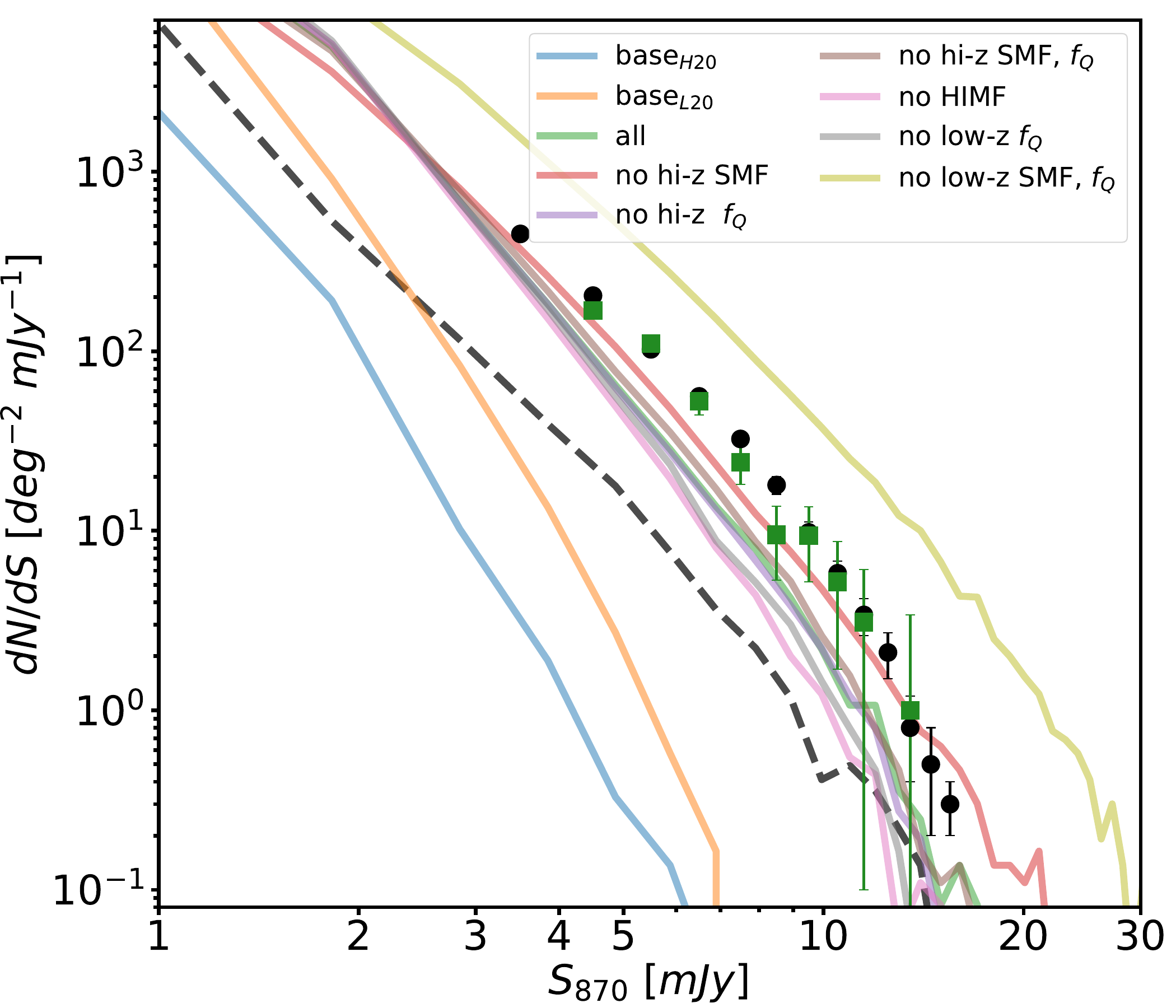}
    \caption{The predicted sub-mm ($870\,\mu\rm{m}$) number counts from a mock catalogue constructed for every configuration listed in Table \ref{tab:configs}, compared to observed number counts derived by \citet{geach17} (black dots) and \citet{stach19} (green squares). As a comparison, the \citet{yo24} number counts from the \citet{henriques15} version of \texttt{L-Galaxies} is included (black dashed line). Most of our calibrated models match the observed $S_{870}$ number counts to within a factor of a few, across an order of magnitude in sub-mm flux density. However, when $n_{\rm SMG}$ is not an observational constraint (`base$_{\rm H20}$' and `base$_{\rm L20}$'), the number counts are critically underpredicted (see orange and blue lines). When neither the low-$z$ SMF nor $f_{\rm{Q}}$ is used as a constraint, sub-mm number counts are overpredicted (see lime green line).}
    \label{fig:number-counts}
\end{figure}

The most accurate way to compare observed differential number counts (number of galaxies per flux-density bin and per unit area) with our model predictions is by constructing mock galaxy catalogs. Following the prescriptions in \citet{yo21} (for sky galaxy positions) and \citet{yo24} (for sub-mm flux densities), we create a $36\,\rm{deg}^2$ mock for the best-fit model of each configuration. Figure \ref{fig:number-counts} shows the predicted $S_{870}$ number counts for all configurations, compared to the observational results from \citet{geach17} and \citet{stach19}. Additionally, we include the $S_{870}$ number counts presented in \citet{yo24}, derived from a mock catalogue (also with a 36 deg$^2$ area) constructed using the \citet{henriques15} version of \texttt{L-Galaxies}.

The first notable result is that including the number density of galaxies with $S_{870}\geq 5.2\,\rm{mJy}$ at $z = 2.8$ as an observational constraint (a single data point) improves the consistency of the predicted $S_{870}$ number counts with observational results. As anticipated from the comparison with direct number density measures in Figure \ref{fig:nsmgs}, configurations `base$_{\rm H20}$' and `base$_{\rm L20}$' severely underpredict the $S_{870}$ number counts, whereas configuration `no low-$z$ SMF, $f_{\rm Q}$' overpredicts them. The remaining configurations show significantly better agreement with the observed number counts compared to the previous version of the model presented in \citet{yo24}, although most still slightly underpredict observed SMG number counts. Among these, we highlight configuration `no hi-$z$ SMF', which nearly matches the observational data. Configurations `all' (calibrated with all observational constraints) and `no hi-$z$ SMF, $f_{\rm Q}$' also demonstrate good agreement with the data, exhibiting similar distributions. In contrast, configuration `no HIMF', which performed better at predicting the low-$z$ quiescent population when the $n_{\rm SMG}$ constraint was included, shows slightly poorer performance in reproducing sub-mm number counts. Nevertheless, even the `no HIMF' configuration shows improved consistency compared to the previous version of the model.

\begin{figure}
    \centering
    \includegraphics[width=\columnwidth]{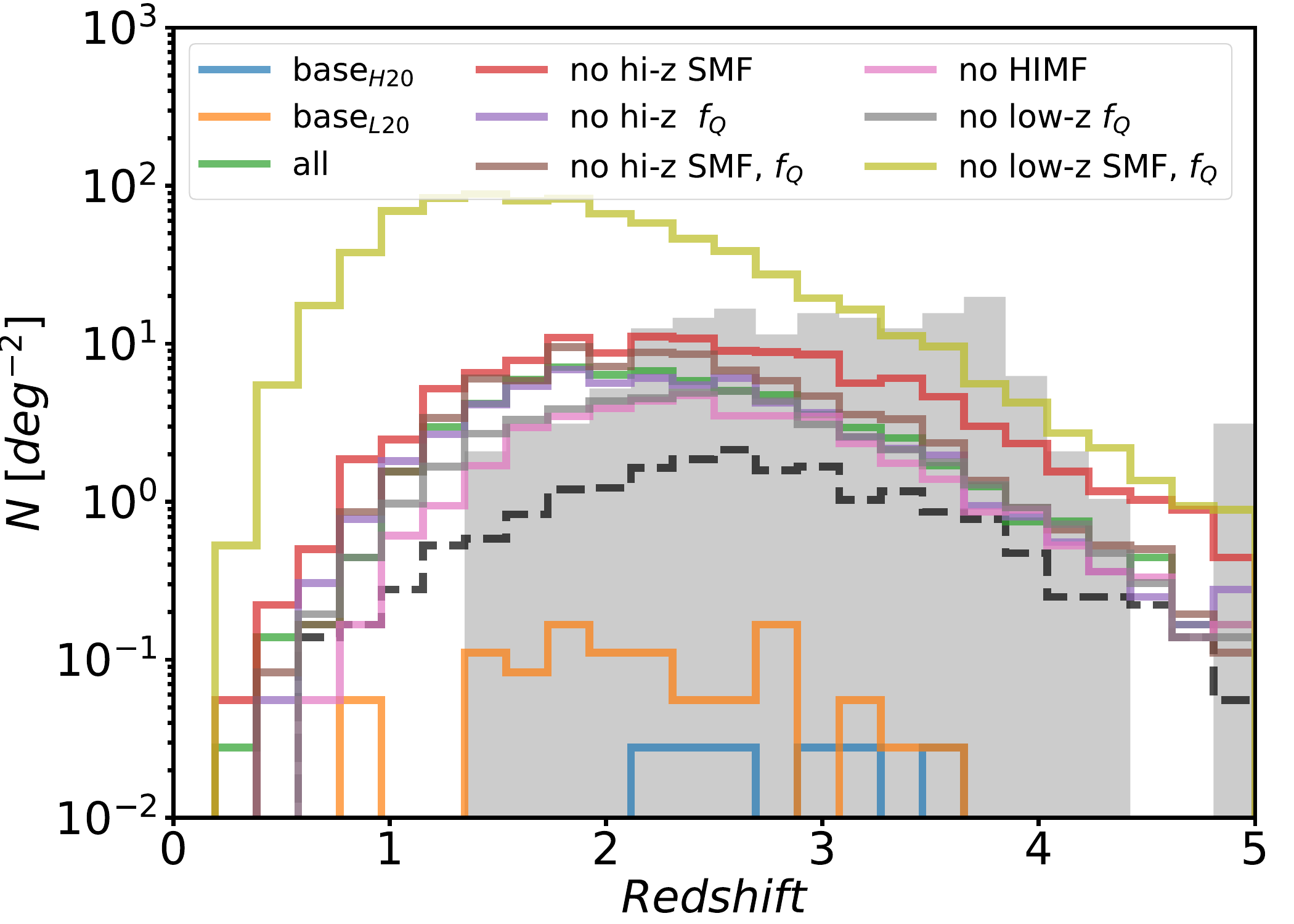}
    \caption{The predicted redshift distribution of bright SMGs ($S_{870}\geq 5.2\,\rm{mJy}$) from a mock catalogue constructed for every configuration listed in Table \ref{tab:configs} (coloured lines), compared to observational data from \citet{dud20} (grey filled histogram). As a comparison, the bright SMG redshift distribution predicted by \citet{yo24} using the \citet{henriques15} version of \texttt{L-Galaxies} is included (black dashed histogram). Most of the configurations present similar redshift distributions.}
    \label{fig:smg_zdist}
\end{figure}

Another valuable comparison with observational data is the redshift distribution of SMGs. We compare our predictions for the number of bright SMGs ($S_{870} \geq 5.2\,\rm{mJy}$) in redshift bins normalized by the sky area, with observational data. Figure \ref{fig:smg_zdist} shows the predictions of the best-fit model for each configuration, compared to AS2UDS data \citep{dud20}. The configuration `no low-$z$ SMF, $f_{\rm Q}$', which overpredicts the sub-mm number counts (Figure \ref{fig:number-counts}), exhibits a peak in the distribution at $z \sim 1.2$, significantly lower than the observed peak at $z \sim 3$. Of the two configurations that critically underpredict the SMG number density, `base$_{\rm H20}$' better matches the shape of the observed redshift distribution compared to `base$_{\rm L20}$'. All other configurations show a similar distribution shape, differing mainly in the number densities across redshift bins, which can be linked to small differences in sub-mm number counts. However, the peak redshift for these models ($z \sim 2$) is also slightly lower than the observed peak. While our best models for predicting sub-mm number counts fail to capture the sharp decrease in galaxy numbers at $z \lesssim 1.4$ , this discrepancy could be due to the small sky area of the observational data and survey selection effects.  

\subsubsection{Number density of Massive Quiescent Galaxies}
\begin{figure}
    \centering
    \includegraphics[width=\columnwidth]{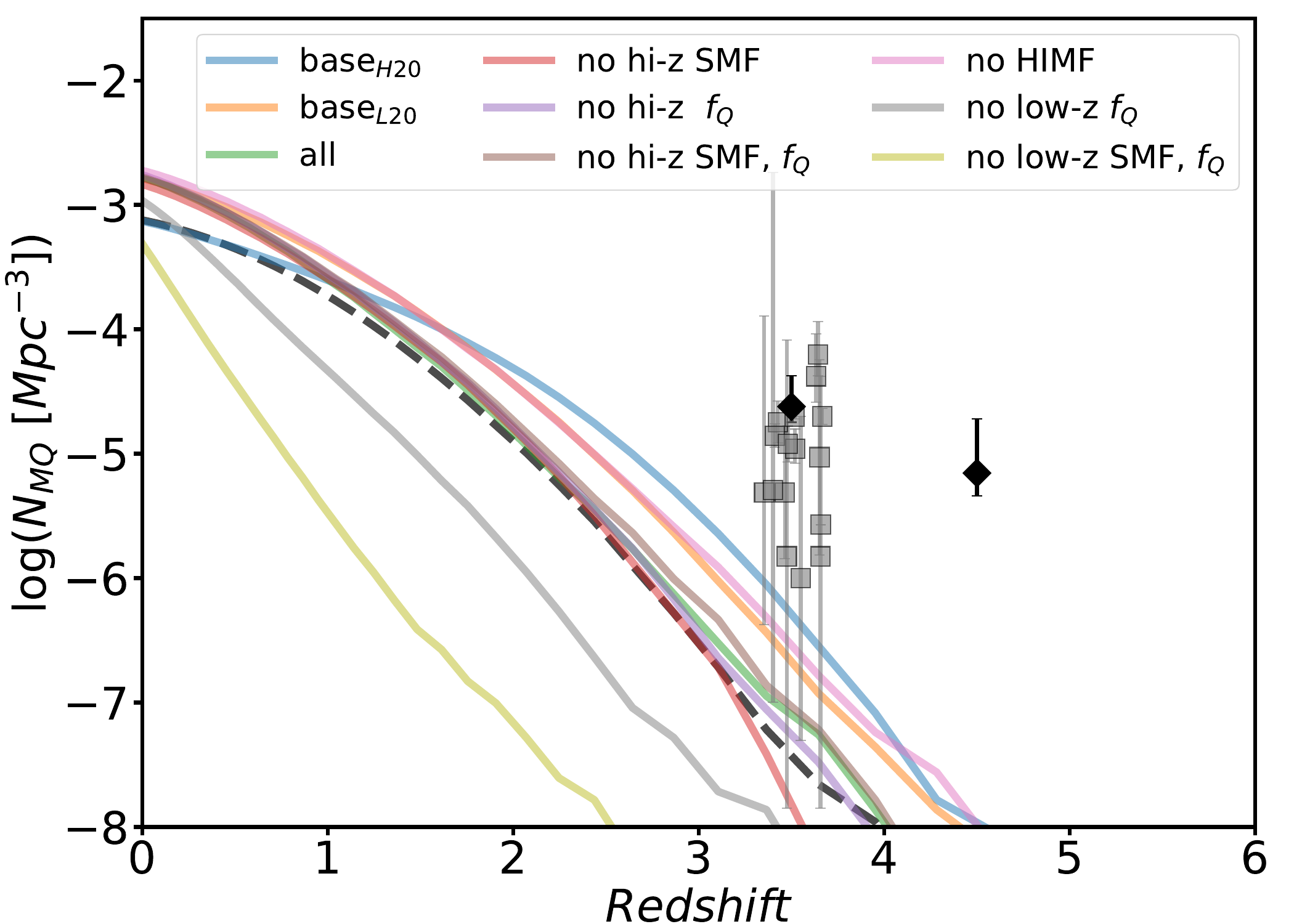}
    \caption{The predicted evolution of the number density of massive ($\log (M_{\star}/\rm{M_{\odot}}) \geq $ 10.6) quiescent (sSFR $< 0.2/t_{\rm obs}(z) $, where $t_{\rm obs}(z)$ is the age of the universe at redshift $z$) galaxies obtained for all configuration listed in Table \ref{tab:configs}. The grey squares are previous observational results compiled by \citet{valentino23}, while the black diamonds are new measurements from \citet{valentino23}. We note that \citet{valentino23} did not detect any MQs at $5<z<6$. As a comparison, the number density from the \citet{henriques15} version of \texttt{L-Galaxies} is included (black dashed line). Except for configuration `no HIMF', when $n_{\rm SMG}$ is not an observational constraint (`base$_{\rm H20}$' and `base$_{\rm L20}$'), the number density of massive quiescent galaxies at high-$z$ is consistent with the lower limits of the observational data. Most of the configurations that match the observed $S_{870}$ number counts underpredict the number density, highlighting the longstanding tension in modelling both populations.}
    \label{fig:nq_ev}
\end{figure}
In this work, we have used the quiescent fraction of massive galaxies at two different redshifts ($z = 0.4$ and/or $z=2.8$) as an input calibration constraint for some configurations. As we show in Figure \ref{fig:fqs}, observational work suggests that the fraction of quiescent galaxies decreases significantly towards high redshift. 
In Figure \ref{fig:nq_ev}, we present the evolution of the number density of massive ($\log (M_{\star}/\rm{M_{\odot}}) \geq  10.6$) quiescent galaxies predicted by our various model configurations. In this figure, we use the QG definition adopted by \citet{carnall20}: sSFR $< 0.2/ t_{\rm obs}(z)$, where $t_{\rm obs}(z)$ is the age of the universe at the redshift $z$. As shown by \citet{carnall20} and  \citet{valentino23}, this definition is virtually equivalent to others commonly used in the literature. Note that this definition is different to that adopted to calibrate the model.
The observational data in Figure \ref{fig:nq_ev} is drawn from \citet{valentino23}, who compiled data from several studies quantifying the number density of massive quiescent galaxies at $3 \leq z \leq 4$. This compilation includes results from \citet{muzzin13, straatman14, davidzon17, schreiber18, merlin19, cecchi19, girelli19, shahidi20, carnall20, weaver23, gould23} and \citet{carnall23}. Note that we add a small scatter ($\Delta z = \pm 0.25$) on the median redshift ($z = 3.5$) of the observational data for visualization purposes. 

The most notable result from Figure \ref{fig:nq_ev} is that none of our configurations can match the median number density of massive quiescent galaxies inferred observationally. Among the configurations, `no low-$z$ SMF, $f_{\rm Q}$', which significantly overpredicts the sub-mm number counts, severely underpredicts the number density of high-redshift massive quiescent galaxies. The configuration with the best performance in this metric is `base$_{\rm H20}$', achieving number densities comparable to the lower bounds of the observations. However, its predictions for sub-mm number counts show the worst agreement with the observational data. Interestingly, the configuration `no hi-$z$ $f_{\rm Q}$' (where $f_{\rm Q}$ at $z = 2.8$ was not used as a constraint) provides better agreement with the observational data than `no low-$z$ $f_{\rm Q}$' (where $f_{\rm Q}$ at $z = 0.4$ was not used as a constraint). This suggests that using $f_{\rm Q}$ at low redshift as a constraint has a greater impact on reproducing the evolution of the quiescent population than applying it only at high redshift. Nevertheless, calibrating the model using the low-redshift $f_{\rm Q}$ does not ensure consistency at higher redshifts, as all configurations struggle to reproduce the observed number density of massive quiescent galaxies.

Overall, the predicted number density of massive quiescent galaxies from configurations where the number density of SMGs was used as a constraint can match only the lower limits of the observational data. The unique exception is the `no HIMF' configuration, whose predictions are similar to those of the base$_{\rm H20}$' and `base$_{\rm L20}$' configurations. Based on Figure \ref{fig:number-counts} and Figure \ref{fig:nq_ev}, our results demonstrate that at high redshift, a higher number density of SMGs corresponds to a lower number density of massive quiescent galaxies, and vice versa. These findings clearly highlight the tension in modelling both the SMG and massive quiescent galaxy populations. Interestingly, the `no HIMF' configuration serves as a `in between the tension' model, producing predictions that do not significantly underrepresent either galaxy population.

\subsubsection{Cosmic Star Formation Rate Density}
\begin{figure}
    \centering
    \includegraphics[width=\columnwidth]{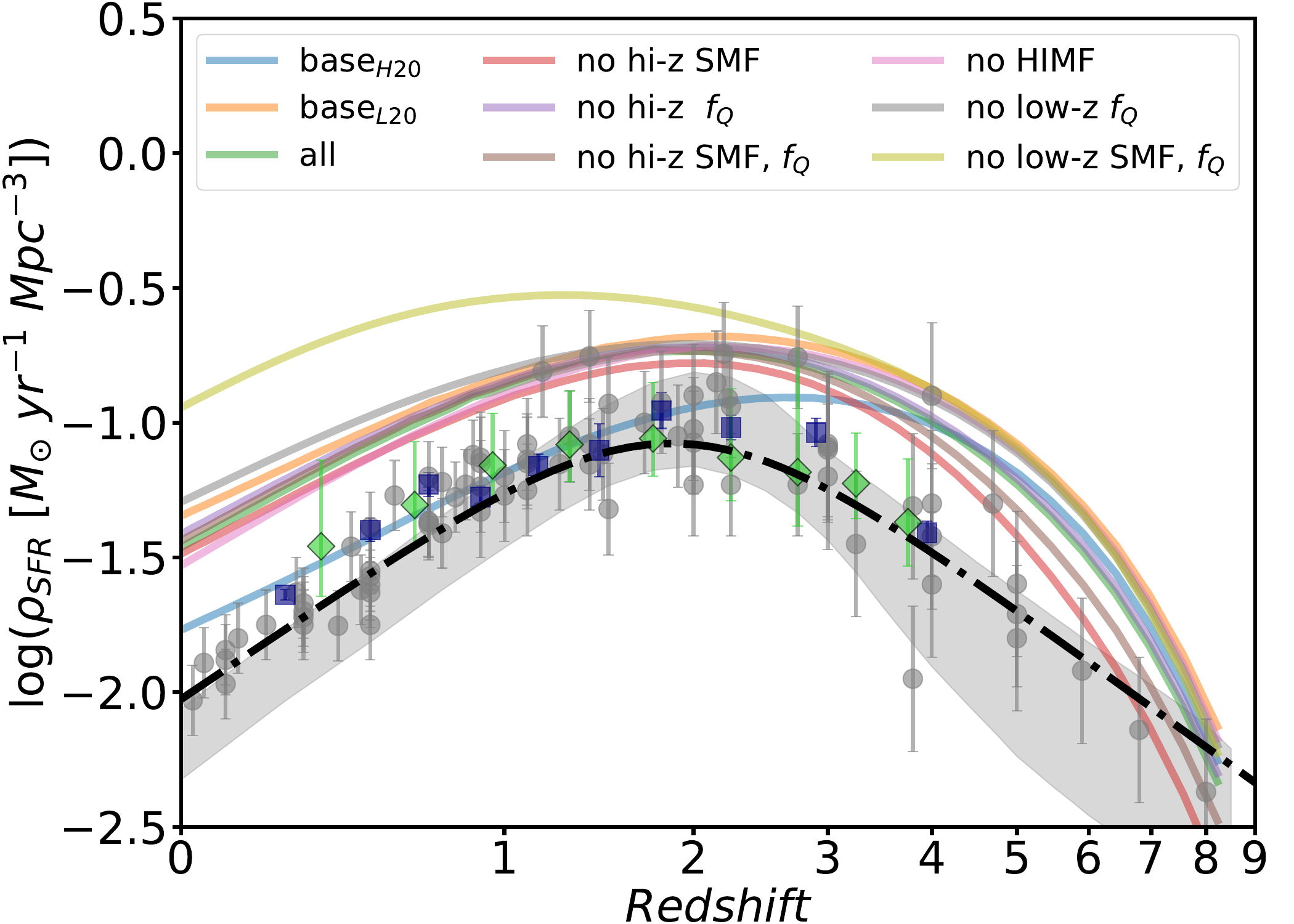}
    \caption{The predicted cosmic star formation rate density (CSFRD) for every configuration listed in Table \ref{tab:configs} compared to \citet{behroozi13} (grey dots), \citet{madau14} function (black dash-dotted line), \citet{leslie20} (green diamonds), \citet{zavala21} (grey area), and \citet{rachel23b} (navy squares) results from observations. Overall, our calibrated models overpredict the star formation rate density, even when $n_{\rm SMG}$ is not used as a constraint. All predicted CSFRDs peak at a similar redshift, which is approximately consistent with observational results, except for configurations `base$_{\rm H20}$' (which peaks earlier) and `no low-$z$ SMF; $f_{\rm Q}$' (which peaks later). }
    \label{fig:sfrd_z}
\end{figure}

One of the most fundamental measures in extragalactic astronomy is the evolving cosmic star formation rate density (CSFRD), which traces the history of star formation and serves as a critical tool for testing galaxy evolution models. Our predictions for the CSFRD are presented in Figure \ref{fig:sfrd_z}, where we compare them to the best-fit function from \citet{madau14} (converted to a \citealt{chabrier03} IMF) and the observational data from \citet{behroozi13, zavala21, leslie20}, and \citet{rachel23b}. All predicted CSFRDs are higher in normalisation than the observational measurements, across all epochs. The CSFRD from configuration `base$_{\rm H20}$' shows better agreement with observational constraints at intermediate redshifts ($0.5 \lesssim z \lesssim 2$), but it overestimates the SFRD both near $z = 0$ and at higher redshifts, with a peak occurring around $z \sim 3$. Notably, the SFRD predicted by configuration `base$_{\rm H20}$' deviates from the results of \citet{henriques20}, particularly in the redshift at which the CSFRD peaks ($z \sim 3$ versus $z \sim 2$ in \citealt{henriques20}). This discrepancy may be attributed to differences in the redshift ranges of the observational constraints used during calibration (e.g. $z = 0.4$ and $z = 2.8$ instead of $z = 0$ and $z = 2$), which can significantly influence the shape of the resulting CSFRD. For instance, configuration `base$_{\rm H20}$' yields high-redshift $f_{\rm Q}$ predictions that are more consistent with observations, whereas \citet{henriques20} tends to underestimate this quantity. Consequently, in order to simultaneously match the high-redshift stellar mass function (SMF) and $f_{\rm Q}$, configuration `base${\rm H20}$' required the peak of star formation to occur earlier than in the \citet{henriques20} model.
On the other hand, configuration `base$_{\rm L20}$' predicts a CSFRD similar to those obtained from configurations where the number density of SMGs is included. This suggests that the difference in shape between the `base$_{\rm H20}$' and the new configurations is primarily driven by our changes to the stellar mass functions used in the calibration. Most configurations predict similar SFRDs below the peak redshift, except for `no low-$z$ SMF, $f_{\rm Q}$', which is approximately $0.5\,\rm{dex}$ higher than the others. Additionally, the CSFRD peak for this configuration occurs at $z \sim 1.2$, aligning with the peak in the SMG redshift distribution for this configuration. In contrast, the other configurations have peaks near $z \sim 2$, consistent with observational data. Slightly larger differences occur at higher redshifts, where the difference between the upper and lower SFRD predictions at $z \sim 8$ is $\sim0.4\,\rm{dex}$.

\subsubsection{SMBH Mass Function}
\begin{figure}
    \centering
    \includegraphics[width=\columnwidth]{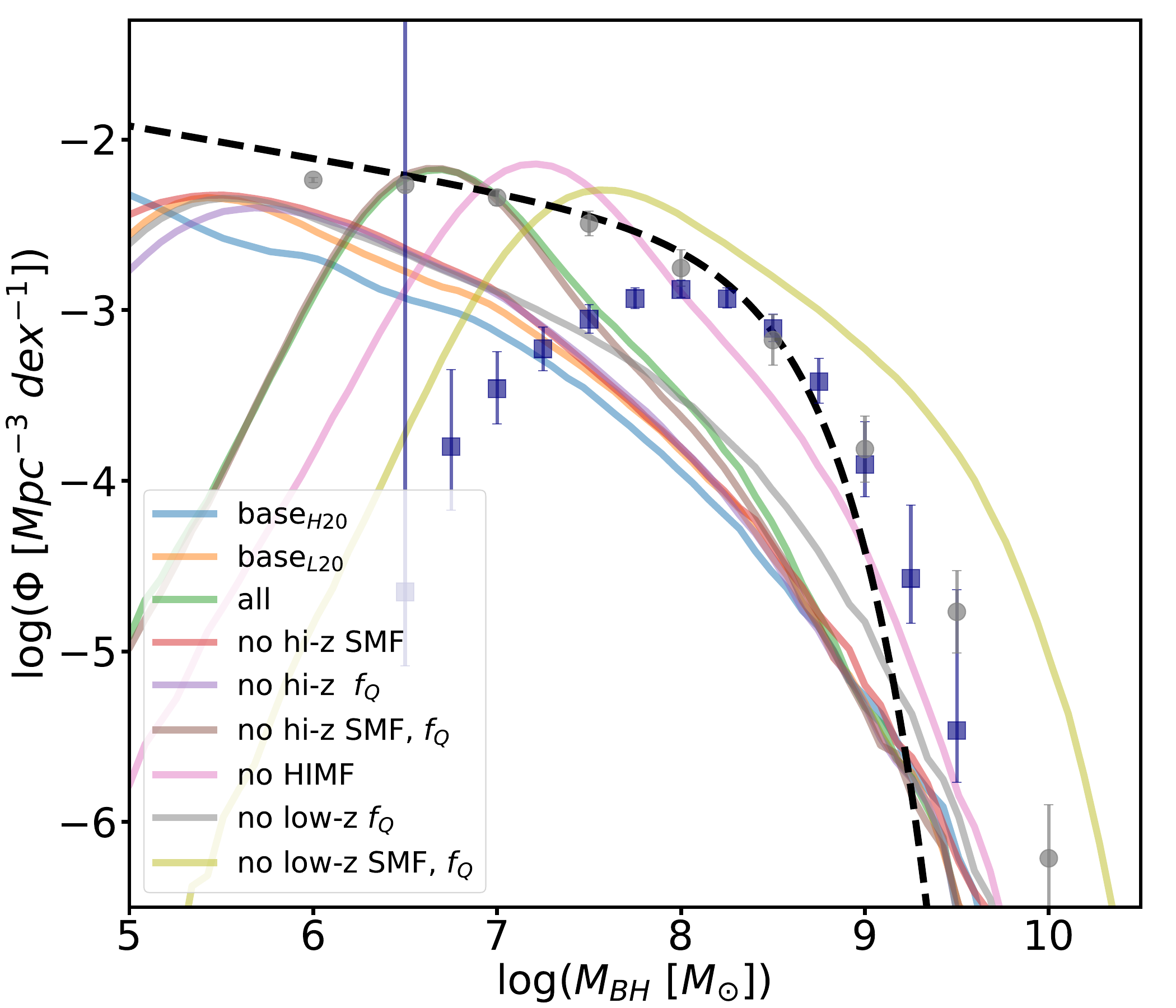}
    \caption{The supermassive black hole mass function at $z = 0$, predicted by each of the configurations listed in Table \ref{tab:configs}. We overplot observational results from \citet{vika09} (navy squares), \citet{tucci17} (dashed black line), and \citet{shankar20} (grey circles) for comparison. The SMBH mass function from configuration `no HIMF' (which is reasonably consistent with the $S_{870}$ number counts and number density of massive quiescent galaxies at high-$z$) is the most consistent with observational data. We identify two main SMBH mass function shapes: peaked distributions and Schechter-like distributions. These shapes are explained by the black hole growth model (see Figure \ref{fig:phys_agn}; top panel).}
    \label{fig:smbh_mf}
\end{figure}

Another important galaxy property is the supermassive black hole (SMBH) mass. In \texttt{L-Galaxies}, the strength of AGN feedback, which injects energy into the hot gas atmosphere, suppressing cooling, depends on the SMBH mass. This process regulates star formation, as only cold gas can form new stars. Note that the SMBH mass function is not used as an observational constraint when calibrating the models.

We present our predictions for the SMBH mass function at $z = 0$ in Figure \ref{fig:smbh_mf}, also comparing to the best-fit function from \citet{tucci17} and observational results from \citet{vika09} and \citet{shankar20}. The predicted SMBH mass functions display two main shapes. The `no low-$z$ SMF, $f_{\rm Q}$', `no HIMF', `no hi-$z$ SMF, $f_{\rm Q}$', and `all' configurations exhibit a clear peak in black hole mass, below which the number densities of SMBHs decrease rapidly. The location of the peak depends on the configuration. For example, the `no low-$z$ SMF, $f_{\rm Q}$' configuration, which overestimates the SMBH mass function, peaks at $\log(M_{\rm BH}/\rm{M_{\odot}}) \sim 7.5$. Configurations `no hi-$z$ SMF, $f_{\rm Q}$' and `all' have similar peaks at $\log(M_{\rm BH}/\rm{M_{\odot}}) \sim 6.7$. In contrast, the `no HIMF' configuration reasonably reproduces the observed SMBH mass function, with its peak at $\log(M_{\rm BH}/\rm{M_{\odot}}) \sim 7.2$. The remaining configurations underpredict the observationally-inferred black hole number densities at intermediate and low masses, with black hole mass functions more similar in shape to SMFs (i.e. not peaked). It is worth noting that a more detailed formalism for the growth of SMBHs in \texttt{L-Galaxies} (a modification of \citealt[][]{henriques15}, as presented in \citealt{izquierdo20}), results in SMBH mass functions that are in good agreement with observations.

\subsection{Physical Interpretation} \label{sec:physcs}

As we have demonstrated, the best-fit models for each configuration produce different predictions for the SMF, $f_{\rm Q}$, HIMF, sub-mm number counts, the evolution of the number density of massive quiescent galaxies, the cosmic star formation rate density, and the SMBH mass function. In this subsection, we present key results that provide insights into the physical processes modelled by each configuration.

\subsubsection{Star Formation}\label{sec:sf}
\begin{figure}
    \centering
    \includegraphics[width=\columnwidth]{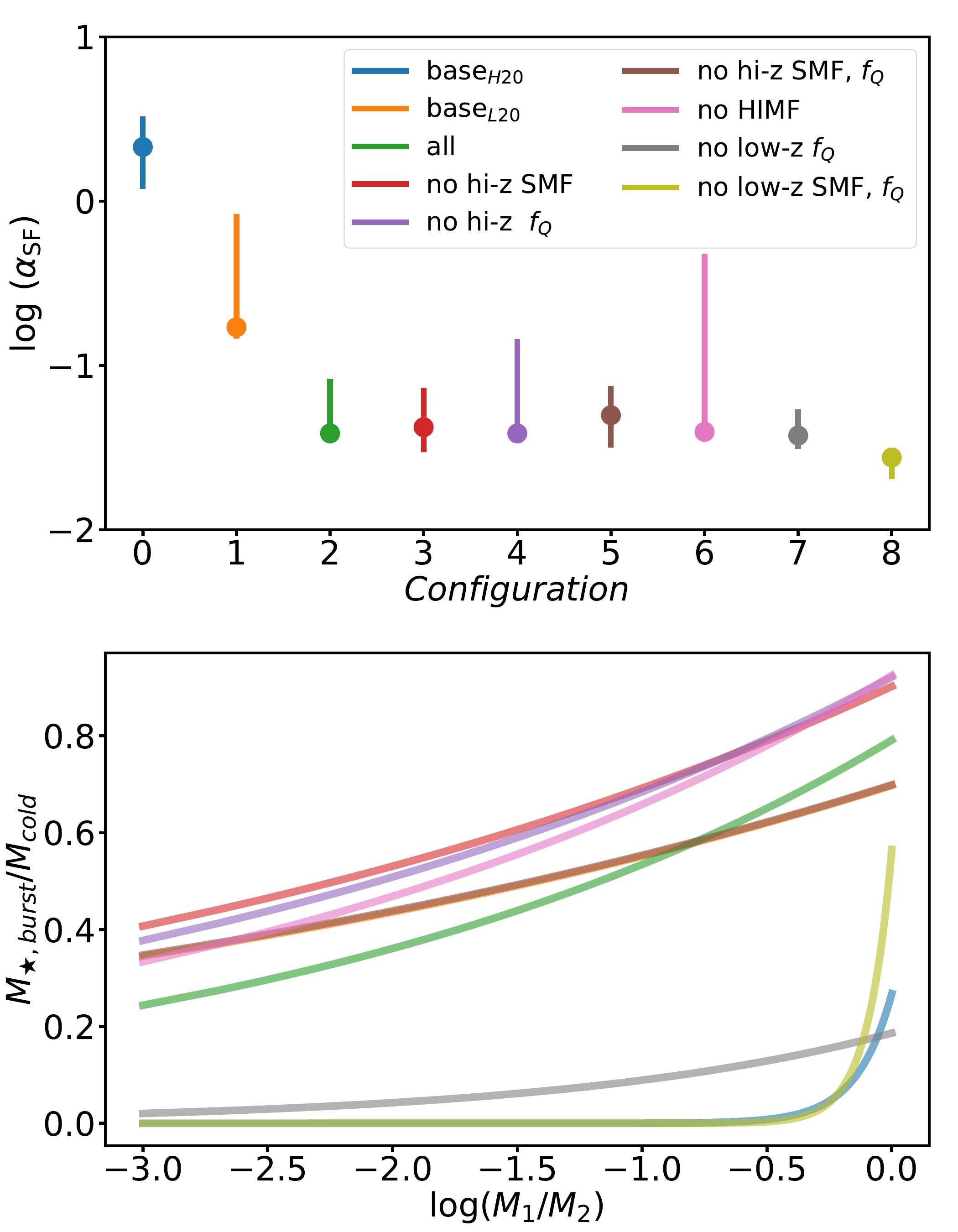}
    \caption{{\it Top:} The best-fit parameter associated with the efficiency in converting H$_2$ into stars (secular star formation). The error bar indicates the $16^{\rm{th}}$ and $84^{\rm{th}}$ percentiles of the final $2,000$ MCMC runs of the $96$ chains. Models calibrated with $n_{\rm SMG}$ present similar and lower efficiency in forming stars from H$_2$ surface density, compared to configurations `base$_{\rm H20}$' and `base$_{\rm L20}$'. {\it Bottom:} The best-fit scaling relation (Equation \ref{eq:sf_burst}) that describes the fraction of cold gas converted into stars driven by mergers as a function of the mass ratio of every configuration. Except for configuration `base$_{\rm H20}$', configurations that critically underpredict the quiescent population (`no low-$z$ $f_{\rm Q}$' and `no low-$z$ SMF, $f_{\rm Q}$') have a low fraction of stars formed in merger-induced starbursts. All other models present an elevated starburst efficiency. This can also deplete the subsequent star formation due to the small amount of remaining cold gas to form new stars.}
    \label{fig:phys_sf}
\end{figure}

In \texttt{L-Galaxies}, there are two physical drivers of star formation: the cold gas surface density (a secular mechanism, implemented as a Kennicutt-Schmidt-type scaling relation) and merger-induced starbursts. We discuss each of these mechanisms, in turn, starting with the secular mechanism. The \citet{henriques20} version of \texttt{L-Galaxies}, which tracks H$_2$ in spatially resolved rings, assumes that the star formation density is proportional to the surface density of H$_2$ \citep{fu13}, with an inverse dependence on the dynamical time. The proportionality constant, $\alpha_{\rm SF}$, is a free parameter in the model. In Figure \ref{fig:phys_sf} (top panel), we show the best-fit $\alpha_{\rm SF}$ for each configuration. We find that the `base$_{\rm H20}$' configuration exhibits the highest efficiency in converting H$_2$ into stars, followed by `base$_{\rm L20}$'. Note that these two models were calibrated without including the number density of SMGs. The remaining configurations, which were calibrated using $n_{\rm SMG}$, display similar values of $\alpha_{\rm SF}$ to each other, with the lowest efficiency seen in the `no low-$z$ SMF, $f_{\rm Q}$' configuration.

The second star formation mechanism is the `collisional starburst' formalism from \citet{somerville01}, which describes the conversion of cold gas into stars triggered by galaxy mergers. This mechanism has two associated free parameters, $\alpha_{\rm SF, burst}$ and $\beta_{\rm SF, burst}$:

\begin{equation}
    M_{\star, {\rm burst}} = \alpha_{\rm SF, burst} \left ( \frac{M_1}{M_2} \right )^{\beta_{\rm SF, burst}} M_{\rm cold},
    \label{eq:sf_burst}
\end{equation}
where $M_1$ and $M_2$ ($M_2 > M_1$) are the total baryonic mass of the two merging galaxies, respectively and $M_{\rm cold}$ is the sum of the cold gas masses (i.e. the sum of HI,  H$_2$, and metals in the ISM). 
We show in Figure \ref{fig:phys_sf} (bottom panel) the scaling relation in Equation \ref{eq:sf_burst} as a function of the galaxy mass ratio for the best-fit models of each configuration. Unlike in the case of secular star formation, configuration `base$_{\rm H20}$' demonstrates the lowest efficiency in converting cold gas into stars during mergers, similar to `no low-$z$ SMF, $f_{\rm Q}$'. In these configurations, no stars form in bursts when the mass ratio of the interacting galaxies is below $\sim$1:3. The efficiency rises quickly for major mergers, reaching $\sim$20\% and $\sim$55\% of the cold gas converted into stars, for the `base$_{\rm H20}$' and `no low-$z$ SMF, $f_{\rm Q}$' models, respectively.  The configuration `no low-$z$ $f_{\rm Q}$' also shows inefficiency in forming stars from mergers, gradually increasing the fraction of gas converted into stars as the mass ratio grows, achieving $\sim$15\% in equal-mass mergers. These three models show significantly different behaviour to the others. The other configurations exhibit strong starburst efficiencies, even at low mass ratios, converting $\sim20-40\%$ of the cold gas into stars during mergers of mass ratio 1:1000 and increasing to $\sim 60-90\%$ in 1:1 mergers. Configurations `base$_{\rm L20}$' and `no hi-$z$ SMF, $f_{\rm Q}$' share identical scalings, as their best-fit parameters related to collisional starbursts are the same. Configurations `no HIMF', `no hi-$z$ SMF', and `no hi-$z$ $f_{\rm Q}$' also show similar scaling relations, with minor differences at low mass ratios. These three configurations are the most efficient in forming stars through this mechanism, achieving up to $\sim90\%$ gas conversion at the highest mass ratios. The `all' configuration shows a similar trend to `no HIMF' but with a lower normalisation. 

Interestingly, configuration `no low-$z$ SMF, $f_{\rm Q}$', which predicts the highest number densities in the SMF at low redshifts and significantly overpredicts sub-mm number counts, is the least efficient in forming stars, in both the secular and merger-driven star formation models. This suggests that the elevated SMF and low fraction of quiescent galaxies seen in that model are due to weaker feedback effects that fail to regulate star formation effectively. We explore feedback in the following subsections. Also of note is the significant difference between the relations fitted for the `base$_{\rm H20}$' and the `base$_{\rm L20}$' models. The star formation in `base$_{\rm H20}$', which was calibrated with lower number densities in the SMFs, is predominantly driven by secular processes, with starbursts only occurring for mergers with mass ratios $\gtrsim 0.5$. In contrast, the `base$_{\rm L20}$' configuration, which was calibrated with the same observables as `base$_{\rm H20}$' but with the SMFs of \cite{leja20}, which have higher normalisation, requires both a high efficiency of secular star formation and significant star formation from merger-induced bursts. This aligns with most configurations that include the number density of SMGs as a constraint.

\subsubsection{Feedback from supernovae and stellar winds}
Especially during the late stages of a star's life, a significant fraction of material and energy is released into the interstellar medium (ISM) through supernovae (SN) and stellar winds. This process is critical for galaxy formation and evolution, as it injects energy that heats the cold gas needed for star formation and enriches the ISM with metals. In \texttt{L-Galaxies}, SN feedback heats the cold gas, transferring it to the hot gas atmosphere. The remaining energy further reheats the hot gas, suppressing cooling and, in some cases, driving outflows. In the \citet{henriques20} version of \texttt{L-Galaxies}, this SN feedback energy is released at the end of a star's life \citep[see][]{yates13}, rather than at the beginning (i.e. the instantaneous recycling approximation) which is commonly assumed in other cosmological simulations. This spreads out the energy (and metals) released per stellar population over time somewhat, particularly for the contribution from SNe-Ia.

The SN feedback model in \texttt{L-Galaxies} is governed by two key efficiencies: reheating cold gas from the disk into the hot gas atmosphere ($\epsilon_{\rm disk}$) and the fraction of the SN energy that is used for ejecting gas from the hot gas atmosphere into the surrounding environment ($\epsilon_{\rm halo}$). Both efficiencies are characterized by the same functional form, parameterized by three free parameters ($\eta_x$, $V_x$, and $\beta_x$):

\begin{equation}
\epsilon_x = \eta_x \times \left[ 0.5 + \left( \frac{V_{\rm max}}{V_x} \right)^{- \beta_x} \right],
\label{eq:sn_feedback}
\end{equation}

\noindent where $x$ represents either the reheat or eject efficiency, and $V_{\rm max}$ is the maximum rotational velocity of the dark matter halo hosting the galaxy. {Since $\epsilon_{\rm halo}$ represents the fraction of SN energy used to eject gas, \texttt{L-Galaxies} does not permit values greater than 1, even though such values are allowed by the functional form. In this sense, the scaling relation may not be the most comprehensive approach for describing the efficiency of SN energy in ejecting hot gas from galaxies. In the original model of \citet{henriques20}, $\epsilon_{\rm halo}$ is saturated at $1$ over the entire $V_{\rm max}$ range.}

\begin{figure}
    \centering
    \includegraphics[width=\columnwidth]{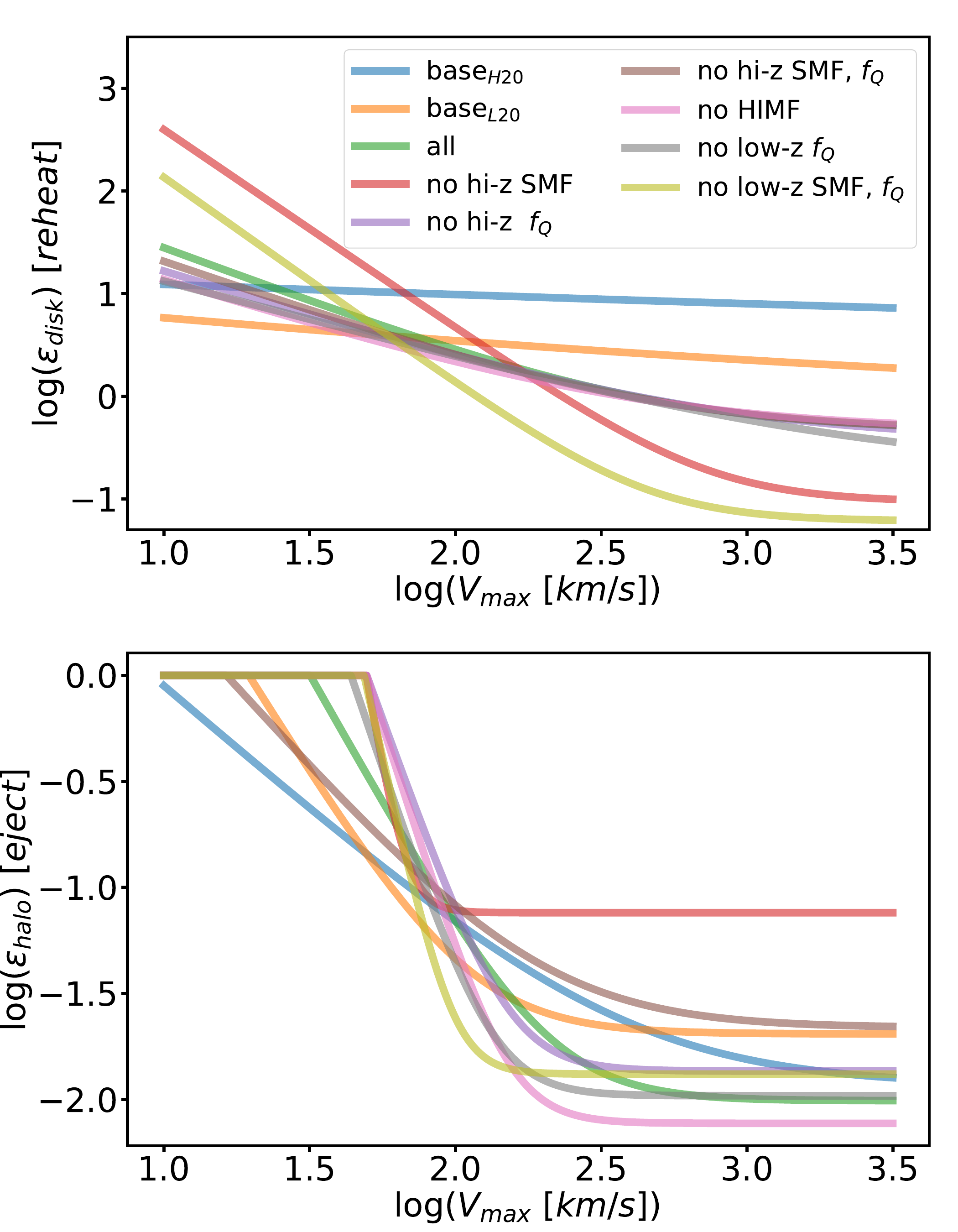}
    \caption{{\it Top:} The best-fit scaling relation (Equation \ref{eq:sn_feedback}) that describes the efficiency of heating the cold gas and reheating the hot gas atmosphere, $\epsilon_{\rm disk}$, as a function of the maximum halo rotational velocity, $V_{\rm max}$ - a proxy of halo mass. The model that best matches the observed $S_{870}$ number counts (`no hi-$z$ SMF') and the model that overpredicts them (`no low-$z$ SMF, $f_{Q}$') both show a heating efficiency that is strongly dependent on halo mass: these models show the least efficient heating in massive halos but the most efficient heating in low-mass halos. On the other hand, the two models that critically underpredict the sub-mm number counts (`base$_{\rm H20}$’ and `base$_{\rm L20}$') present a weak dependence in $V_{\rm max}$, having the strongest efficiency of heating cold gas for high-mass halos. {\it Bottom:} The best-fit scaling relation (Equation \ref{eq:sn_feedback}) that describes the fraction of available SN energy to eject gas (in outflows) from the galaxy's hot gas atmosphere, $\epsilon_{\rm halo}$, as a function of the maximum halo rotational velocity, $V_{\rm max}$. The nine models have similar efficiency for ejecting gas in intermediate and high-mass systems, except for configuration `no hi-$z$ SMF'. This could be the reason for this configuration predicting the lowest SFRD at higher redshifts.} 
    \label{fig:phys_sn}
\end{figure}

We present the scaling relations for reheating and ejecting gas as a function of $V_{\rm max}$ (a proxy for halo mass) for the best-fit models of each configuration in the top and bottom panels of Figure \ref{fig:phys_sn}. The configurations predict three distinct scaling relation shapes for the reheating mechanism. Configurations `base$_{\rm H20}$’ and `base$_{\rm L20}$' exhibit an almost linear relation, with the weakest dependency on $V_{\rm max}$. Despite this, their efficiencies for reheating cold gas are the highest among all configurations, for halos with $\log (V_{\rm max}) \gtrsim 1.9$ (`base$_{\rm H20}$') and $\log (V_{\rm max}) \gtrsim 2.1$ (`base$_{\rm L20}$'). The higher efficiency in heating gas for massive halos seems to be important for reproducing the massive quiescent population. In contrast, configurations `no low-$z$ SMF, $f_{\rm Q}$’ and `no hi-$z$ SMF' show a strong inverse dependency on $V_{\rm max}$, differing only by an almost constant offset. These models are the most efficient at reheating cold gas in low-mass halos but the least efficient in massive halos. As a result, the `no low-$z$ SMF, $f_{\rm Q}$' configuration overpredicts the SMFs and the CSFRD at low redshift; this is due to the greater availability of gas reservoirs allowed by the weak SN and AGN feedback (see also next subsection) rather than an increased efficiency in forming stars.  
The remaining configurations display strong similarities in their scaling relations for reheating cold gas through SN feedback, with minor variations in specific parameter values.

The second scaling relation (Figure \ref{fig:phys_sn}, bottom panel) describes the fraction of SN energy used for ejecting gas from the hot gas atmosphere. At high $V_{\rm max}$ ($\log (V_{\rm max} / {\rm km \, s}^{-1})\gtrsim2$), most configurations exhibit similar efficiencies in massive halos, spanning between $\log(\epsilon_{\rm halo}) \sim -2$ and $\log(\epsilon_{\rm halo}) \sim -1.5$. Notice that the fraction of SN feedback is saturated ($\epsilon_{\rm halo} = 1$) in most configurations at $V_{\rm max} \lesssim 50 \, \mathrm{km/s}$. This velocity is typical for dark matter halos with $M_{\rm vir} \lesssim 10^{11}\,\rm{M_{\odot}}$. The `no hi-$z$ SMF’ configuration has the highest ejection efficiency across almost the entire $V_{\rm max}$ range. The particularly strong feedback in low and intermediate mass halos, which are more prevalent in the early universe, provides a physical explanation for why this configuration predicts the lowest SFRD at high redshifts. 
On the other hand, configuration `no low-$z$ SMF, $f_{\rm Q}$' also presents high efficiency in ejecting and reheating cold gas in low-mass systems, as well as the lowest efficiency in forming stars (both secular and merger-driven; see Figure \ref{fig:phys_sf}). However, the predicted SFRD for this configuration is comparable to that of other configurations at high redshift and even higher at low redshift. This is driven by the larger number of intermediate mass ($M_{\star}\sim10^{8.5}-10^{10}\,\rm{M_{\odot}}$) star-forming galaxies in that model, as described in Section \ref{sec:sf}.

\subsubsection{AGN Feedback}
\begin{figure}
    \centering
    \includegraphics[width=\columnwidth]{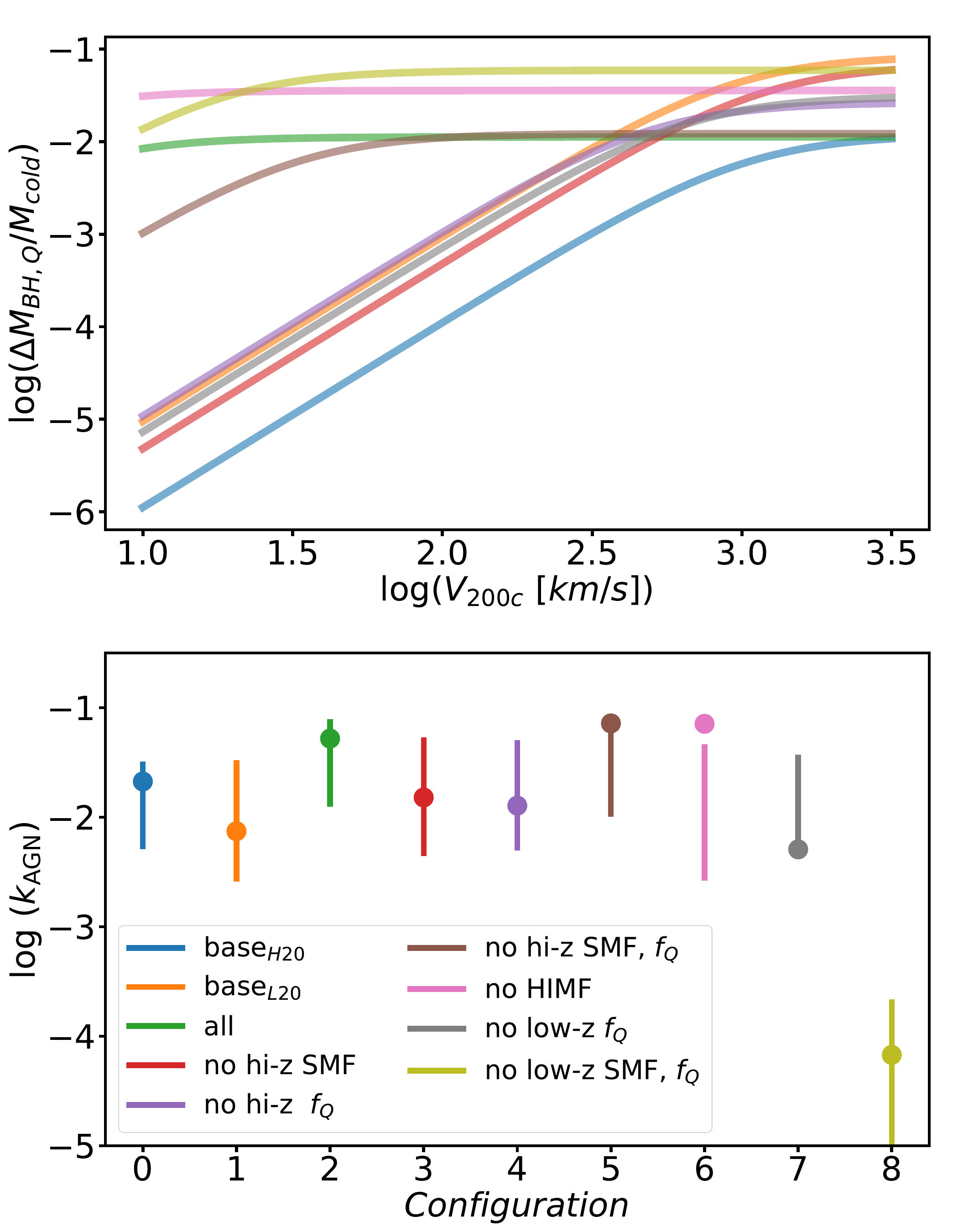}
    \caption{{\it Top:} The best-fit scaling relation (Equation \ref{eq:quasar_mode}) that describes the black hole growth from cold gas accretion in merger events (assuming $M_{\rm sat}/M_{\rm cen} = 1$, for the purpose of illustration) as a function of $V_{200c}$ - a proxy of halo mass. Models that predict a peak-shaped SMBH mass function at $z=0$, including the model that better matches the observational results (`no HIMF'), present an almost halo mass-independent cold gas accretion. The SMF-like SMBH mass function is driven by the strong dependence of the accreted cold gas by the SMBH on $V_{200c}$. {\it Bottom:} The best-fitting AGN efficiency parameter, $k_{\rm AGN}$, for every configuration. As in the top panel of Figure \ref{fig:phys_sf}, the error bar indicates the $16^{\rm{th}}$ and $84^{\rm{th}}$ percentiles of the final $2,000$ MCMC runs of the $96$ chains. All models, except configuration `no low-$z$ SMF, $f_{\rm Q}$', have similar AGN efficiency in injecting energy to the hot gas, reducing the cooling rate. Thus, the AGN feedback across the models becomes almost entirely dependent on the SMBH mass, being underestimated due to the underpredictions of the SMBH mass function, except for configuration `no HIMF'. Although configuration `no low-$z$ SMF, $f_{\rm Q}$' overpredicts the SMBH mass function, the AGN feedback, in this case, is less impactful due to the low AGN efficiency, $k_{\rm AGN}$. }
    \label{fig:phys_agn}
\end{figure}

As shown in Figure \ref{fig:phys_sn}, SN and stellar wind feedback are less effective at heating or ejecting cold gas in massive systems. Consequently, active galactic nucleus (AGN) feedback becomes an essential mechanism for energy release in high-mass galaxies. The 2020 version of the \texttt{L-Galaxies} model follows the \citet{croton06} framework to implement AGN feedback, which operates in two modes: quasar and radio modes. The quasar mode, while not directly associated with energy release (other than from the associated merger-induced starburst), describes the growth of SMBHs, whereas the radio mode injects energy into the hot gas atmosphere, suppressing cooling. In the \citet{croton06} framework, the primary mechanism for SMBH growth is the quasar mode. During galaxy mergers, a fraction of the cold gas is either accreted by the pre-existing SMBH or contributes to forming a new SMBH. The amount of cold gas accreted is given by:

\begin{equation} 
\Delta M_{\rm BH, Q} = \frac{f_{\rm BH} (M_{\rm sat}/ M_{\rm cen}) M_{\rm cold}}{1 + (V_{\rm BH}/V_{200c})^2},
\label{eq:quasar_mode} 
\end{equation}
where $M_{\rm cen}$ and $M_{\rm sat}$ are the baryonic masses of the central and satellite galaxies, respectively, and $M_{\rm cold}$ is the sum of their cold gas components. $V_{200c}$ represents the virial velocity of the host dark matter halo, while $f_{\rm BH}$ and $V_{\rm BH}$ are free parameters in the model.

We present the fraction of cold gas accreted by SMBHs, $\Delta M_{\rm BH, Q}/M_{\rm cold}$, as a function of $V_{200c}$, for each best-fit model configuration in the top panel of Figure \ref{fig:phys_agn}. For illustration, we assume $M_{\rm sat}/M_{\rm cen} = 1$, although this ratio is generally $\leq 1$. The configurations predict two main trends in the dependence of $\Delta M_{\rm BH, Q}/M_{\rm cold}$ on $V_{200c}$. The configurations labeled `no low-$z$ SMF, $f_{\rm Q}$,’ `no HIMF,’ `all,’ and `no hi-$z$ SMF, $f_{\rm Q}$’ exhibit high efficiency in accreting cold gas during mergers, with minimal dependency on dark matter halo mass. 
These four configurations also display a peak-shaped SMBH mass function at $z=0$ (see Figure \ref{fig:smbh_mf}). 
Interestingly, all the configurations with a peak-shaped SMBH mass function at $z = 0$ also show a near-independence of the fraction of accreted cold gas by the SMBHs on dark matter halo mass.

The other configurations exhibit a strong dependence of the fraction of cold gas that is accreted onto the black hole on halo mass. Interestingly, while the `no HIMF’ and `no low-$z$ SMF, $f_{\rm Q}$’ configurations follow similar SMBH growth scalings, their predicted SMBH mass functions differ significantly. This discrepancy may arise from the larger cold gas reservoirs allowed by the weak SN (and AGN; see below) feedback in the `no low-$z$ SMF, $f_{\rm Q}$’ configuration. The remaining configurations exhibit comparable scaling relations for the SMBH growth and, consequently, similar SMBH mass functions. The scaling relation for these configurations strongly depends on $V_{200c}$. Although the `base$_{\rm H20}$’ model is approximately $1\,\rm{dex}$ less efficient than other configurations (with dependency on $V_{\rm 200c}$), the resulting SMBH mass function at $z = 0$ shows is still similar, differing only by $\sim 0.1$ dex.

The second SMBH-related process in \texttt{L-Galaxies} is the radio mode, which explicitly models AGN feedback. This mechanism involves gas accretion from the hot atmosphere onto the SMBH and the subsequent injection of energy into the hot gas, suppressing cooling. The \citet{croton06} formalism defines the SMBH accretion rate as a function of the hot gas and SMBH masses, with a free parameter, $k_{\rm AGN}$, representing the efficiency of hot gas accretion and energy injection. The best-fit values of $k_{\rm AGN}$ for each configuration are shown in Figure \ref{fig:phys_agn} (bottom panel). Overall, $k_{\rm AGN}$ is consistent across most configurations, except for `no low-$z$ SMF, $f_{\rm Q}$,’ where it is approximately three orders of magnitude lower. Despite predicting rapid SMBH growth, the low AGN efficiency in this configuration renders the feedback ineffective at quenching star formation. Configurations with the highest $k_{\rm AGN}$ efficiency also predict a peak-shaped SMBH mass function and minimal dependency of cold gas accretion on dark matter halo mass; this highlights AGN feedback as a critical quenching mechanism in these models. Interestingly, the similar $k_{\rm AGN}$ values across most configurations, despite differences in the number density of massive quiescent galaxies (Figure \ref{fig:nq_ev}), suggest that other mechanisms — such as strong SN feedback (as seen in the `base$_{\rm H20}$’ and `base$_{\rm L20}$’ configurations at high $V_{\rm{max}}$) — play a significant role in generating this population.

\subsection{Model degeneracy}

\begin{figure}
    \centering
    \includegraphics[width=\columnwidth]{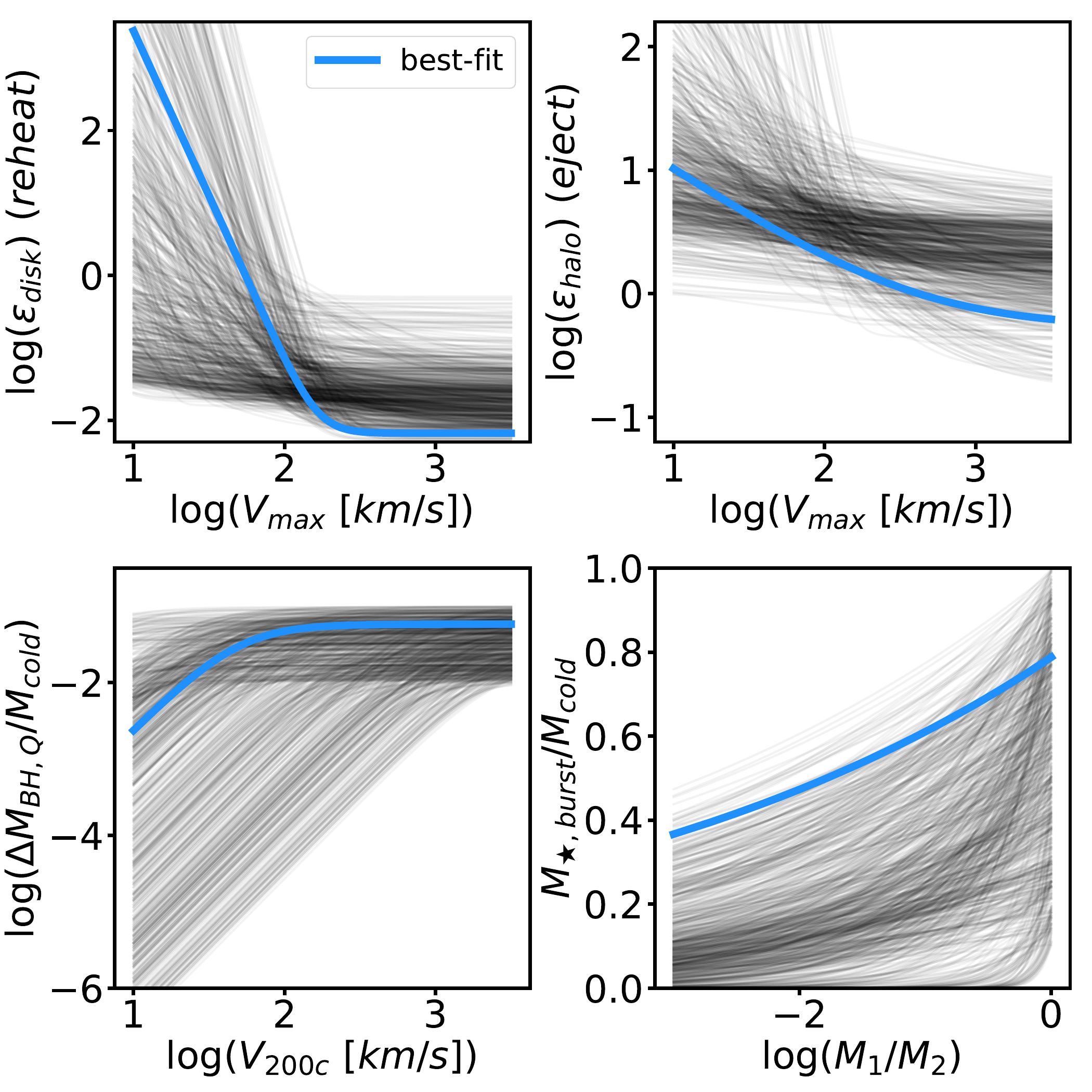}
    \caption{Physical scaling relations that describe the SN feedback, $\epsilon_{\rm disk}$ and $\epsilon_{\rm disk}$ (top panels; Equation \ref{eq:sn_feedback}), SMBH growth (bottom left panel; Equation \ref{eq:quasar_mode}), and stellar mass formed in merger-induced starbursts (bottom right; Equation \ref{eq:sf_burst}) for all sets of free parameters with total likelihood within $1\,\rm{dex}$ of the best-fit (transparent black lines) for configuration `no HIMF'. These results evidence the high level of model degeneracy, where the models with similar likelihood, in some cases, range many orders of magnitude. In most cases, the scaling relations from the configuration best-fit do not occupy the most populated area drawn from similar likelihood models.}
    \label{fig:deg_model}
\end{figure}

The best-fit model for each configuration corresponds to the set of free parameters that yield the highest likelihood. However, given the complex hyperparameter space of \texttt{L-Galaxies}, degeneracies are expected. In this section, we analyze how the physical scaling relations presented in the previous section vary across models with similar likelihoods.

Figure \ref{fig:deg_model} illustrates four physical scaling relations for parameter sets with likelihoods within $1\,\rm{dex}$ of the maximum likelihood, based on the final $2,000$ MCMC runs across the $96$ chains (see Section \ref{sec:run}) of the `no HIMF' configuration. 
The physical scalings shown in Figure \ref{fig:deg_model} are: the efficiency of reheating, $\epsilon_{\rm disk}$ (top left panel; corresponding to the top panel of Figure \ref{fig:phys_sn}), the efficiency of ejecting gas, $\epsilon_{\rm halo}$ (top right panel; corresponding to the bottom panel of Figure \ref{fig:phys_sn}), SMBH growth from cold gas, $\Delta M_{\rm BH, Q}/M_{\rm cold}$ (bottom left panel; corresponding to the top panel of Figure \ref{fig:phys_agn}), and the fraction of cold gas converted into stars during merger-induced starbursts, $M_{\star, {\rm burst}} / M_{\rm cold}$ (bottom right panel; corresponding to the bottom panel of Figure \ref{fig:phys_sf}).

As shown in Figure \ref{fig:deg_model}, the physical scaling relations for {\it good} likelihood models exhibit significant scatter, spanning approximately two dex across the ranges of $V_{\rm max}$, $V_{\rm 200c}$, and $M_1 / M_2$. Also, the physical scalings derived from the best-fit model of the `no HIMF' configuration generally do not coincide with the most densely populated regions (representing models with similar scalings), except in the case of SMBH growth from cold gas. These findings highlight the high level of degeneracy in the galaxy formation model when calibrated against the observational constraints of the `no HIMF' configuration (as is also the case for the other configurations). This suggests that the physical insights that can be gained from such fitted parameters may be limited. 

\section{Discussion} \label{sec:discussion}

The tension between observations and theoretical models in simultaneously modelling dusty star-forming galaxies (DSFGs) and high-redshift massive quiescent galaxies (MQs) remains unresolved. Many models struggle to reproduce even one of these extreme populations, particularly the MQ population, as shown by \citet{lagos25}. However, given the current approach to setting the free parameters of astrophysical processes — often done manually — it is not entirely evident that galaxy formation models fundamentally fail to reproduce these populations. In this work, we test the ability of the \texttt{L-Galaxies} semi-analytic model to address this tension by performing a robust calibration designed to reproduce observed populations of both DSFGs and MQs. We chose \texttt{L-Galaxies} for this study because its MCMC-based calibration framework allows us to explore different sets of observational constraints systematically. Our results yield a model that represents progress towards resolving this discrepancy, though some limitations remain. The key findings of this work are discussed in this section.

\subsection{The impact of varying observational constraints}
The adopted calibration framework was designed to explore how the calibrated model (defined by the set of tuned parameters) varies given different sets of observational constraints (see Table \ref{tab:configs}). For instance, the configurations named `base$_{\rm H20}$' and `base$_{\rm L20}$' use essentially the same sets of constraints, differing only in the source of the dataset used to constrain stellar mass functions and massive quiescent galaxy fractions. The `base$_{\rm H20}$' configuration uses the same data as \citet{henriques20} (a compilation from the literature), whereas `base$_{\rm L20}$' was calibrated with the stellar mass functions (SMFs) and quiescent fractions ($f_{\rm Q}$) from \citet{leja20} and \citet{leja22}, respectively. The key difference between these datasets lies in the SMFs, with \citet{leja20} predicting systematically higher number densities across the stellar mass range, particularly at the massive end. As expected, this leads to differences in the SMFs predicted by the calibrated models (Figure \ref{fig:smfs}) and significantly impacts the predicted cosmic star formation rate density (Figure \ref{fig:sfrd_z}), with a discrepancy of approximately $0.5\,\rm{dex}$ at low redshift. While both configurations reproduce the quiescent population reasonably well and in a similar manner (Figure \ref{fig:fqs}), they severely underpredict the sub-millimetre number counts (Figure \ref{fig:number-counts}). In these configurations, number density of SMGs ($n_{\rm SMG}$) was not included as an observational constraint. Our results motivate the incorporation of $n_{\rm SMG}$ as a constraint.\\
\indent In the remaining configurations, we address this issue by including sub-mm number densities within a systematic MCMC calibration framework. This work represents the first time such a robustly-calibrated semi-analytic model has included SMGs as a constraint. We calibrated these configurations using the number density of bright ($S_{870} \geq 5.2\,\rm{mJy}$) SMGs at $z = 2.8$ to capture the observed sub-mm number counts, while also fitting for SMFs and $f_{\rm Q}$ across different redshifts. Among all configurations, only the `no hi-$z$ SMF' successfully reproduces the observed $S_{870}$ number counts, whereas the others tend to slightly underpredict them. On the other hand, omitting low-redshift constraints (configuration `no low-$z$ SMF, $f_{\rm Q}$') leads to an {\it{overprediction}} of the sub-millimetre number counts. The main differences in the predictions arise in the quiescent population, with all configurations underrepresenting these galaxies to some degree. When $f_{\rm Q}$ at low redshift is not included as an observational constraint, the best-fit model significantly underestimates the number of quiescent galaxies (configurations `no low-$z$ SMF, $f_{\rm Q}$' and `no low-$z$ $f_{\rm Q}$'). Conversely, when the high-redshift SMF and $f_{\rm Q}$ are not used as constraints, the resulting models are similar (configurations `no hi-$z$ SMF', `no hi-$z$ $f_{\rm Q}$', and `no hi-$z$ SMF, $f_{\rm Q}$').

Interestingly, when the high-redshift SMF is not used as a constraint (configuration `no hi-$z$ SMF'), the best-fit model most accurately reproduces the observed number counts but performs the worst in predicting the number density of massive quiescent galaxies at high redshift — even more so than when the high-redshift $f_{\rm Q}$ is not included. Additionally, we find that when both SMF and $f_{\rm Q}$ are used simultaneously as constraints (configuration `all'), the predicted galaxy properties remain statistically similar to those obtained when neither constraint is applied (configuration `no hi-$z$ SMF, $f_{\rm Q}$').

Finally, although the neutral hydrogen mass function (at $z = 0$) was not included as a constraint in the `no HIMF' configuration, the model still successfully reproduces it. Among the best-performing models, this particular configuration is slightly weaker in predicting both the sub-millimetre number counts and the number density of massive quiescent galaxies at high redshift. Nevertheless, it still achieves a reasonable agreement for both, making it the most successful model overall in simultaneously reproducing DSFGs and MQs.

\subsection{Matching the sub-mm number counts}

Matching the sub-millimetre number counts without invoking an IMF modification remains a challenge for many of the most widely used cosmological simulations, such as \texttt{EAGLE} \citep{cowley19}, \texttt{IllustrisTNG} \citep{chris21}, and \texttt{L-Galaxies} \citep{yo24}, among others. Only a few simulations have been able to closely reproduce the observed number counts, including \texttt{Illustris} \citep{chris21}, \texttt{SIMBA} \citep{lovell21}, \texttt{FLAMINGO} \citep{kumar25}, and the \texttt{SHARK} (v1.0) \citep{lagos19} semi-analytic model.

In this study, we demonstrated that incorporating the number density of bright SMGs at a single redshift as a constraint significantly improves model predictions of the sub-mm number counts. Indeed, all configurations that included this constraint successfully matched (or closely matched) the observational data, across an order of magnitude in sub-mm flux density. Among these, configuration `no hi-$z$ SMF' provides the best predictions for the sub-mm number counts, even at the bright end. These new calibrated models present an opportunity for future theoretical studies of bright SMGs.

Here, we explore the key characteristics that favour good matches to the observed SMG population, based on the best-fitting parameters presented in Section \ref{sec:physcs}. 
The configuration that best-reproduces the observed number counts (`no hi-$z$ SMF') has similar best-fitting parameters to most of the other models, for parameters relating to star formation (both secular and merger-driven), SMBH growth, and AGN feedback efficiency.
The main difference lies in the supernova feedback model, specifically in the scaling relation for (re)heating gas (Figure \ref{fig:phys_sn}, top panel). The two configurations that predict the highest number of SMGs (`no hi-$z$ SMF' and `no low-$z$ SMF, $f_{Q}$') exhibit similar functional forms for this efficiency, which plays a crucial role in suppressing star formation. The best-fit scaling relations indicate a strong inverse dependence of the (re)heating efficiency on $V_{\rm max}$ (proxy of dark matter halo mass). For these configurations, the efficiency is highest in low-mass systems and lowest in high-mass haloes. As a result, more cold gas remains available for star formation in massive galaxies (the mass range of most SMGs) compared to the other models, whether this star formation proceeds through secular or merger-induced star formation. 

Given the high star formation efficiencies involved in these two calibrated models for high stellar mass galaxies, AGN feedback would be required for effective quenching. However, the `no hi-$z$ SMF' configuration actually {\it{underpredicts}} the SMBH mass function at $z = 0$ (Figure \ref{fig:smbh_mf}), indicating that SMBH growth in this model is lower than required by observations. Consequently, AGN feedback is less effective than required to reproduce the quiescent population, and number densities of massive quiescent galaxies in this model are significantly lower than observed. On the other hand, the `no low-$z$ SMF, $f_{\rm Q}$' configuration exhibits a similar SN feedback scaling relation to no hi-$z$ SMF' but critically overpredicts the sub-mm number counts. In this case, unlike in the other models, the SMBHs grow beyond the observed values, yet the AGN feedback efficiency (radio mode) is approximately two orders of magnitude lower. As a result, the `no low-$z$ SMF, $f_{\rm Q}$' configuration overpredicts the stellar mass functions, sub-mm number counts, and cosmic star formation rate density, as there is no strong regulatory mechanism to suppress star formation in intermediate and massive galaxies. These cases highlight the persistent difficulty of reproducing both SMG and MQ populations.  

\subsection{Reproducing the massive quiescent population at high-$z$}
Galaxy formation models fail to reproduce the high number density of massive quiescent galaxies at $z \gtrsim$ 3 from observations, as shown in \citet{merlin19}, \citet{szpila25}, \citet{vani25}, and \citet{lagos25}, among others; this is of considerable interest given the increasing numbers of such galaxies being characterised by JWST. Indeed, this issue also happens for \texttt{L-Galaxies}. However, it is important to note that there are large discrepancies between current observational estimates of the number density, with limits ranging $\sim2\,\rm{dex}$ \citep{valentino23}. This is, at least in part, due to the different methods to estimate galaxy properties, selection criteria, and the available data used in those works. Cosmic variance and the difficulties of estimating number densities from extremely small samples also pose a challenge for these works. For instance, \citet{alberts24} obtained comparable number densities of MQs to \citet{carnall23} and \citet{valentino23}, despite adopting a $\sim1\,\rm{dex}$ lower stellar mass cut and explicitly studying an overdense region. 

Most of our re-calibrated models underestimate both the fraction and number density of quiescent galaxies (Figures \ref{fig:fqs} and \ref{fig:nq_ev}, respectively), especially at high redshift.\footnote{One possible solution is more accurate modelling of the properties of orphan galaxies — galaxies whose host dark matter halos have been fully accreted or disrupted by more massive systems — in SAMs, as demonstrated by \citet{harrold24} for the \citet{henriques15} version of \texttt{L-Galaxies}. This approach significantly improves the consistency with the observed quiescent galaxy stellar mass function.} Among our configurations, only three are in reasonable agreement with the lower limits of the observed number density of MQs, comparable with some models presented by \citet{lagos25}. These configurations are `base$_{\rm H20}$', `base$_{\rm L20}$', and `no HIMF', the first two of which provide the worst matches to the sub-mm number counts. In contrast, the configuration `no HIMF' closely matches the sub-mm number counts \citep[similar to the results of][who studied SMGs in the \texttt{SIMBA} simulation]{lovell21} and hence provides a promising avenue for future work.

We identify two different combinations of physical mechanisms that act to quench galaxies in these three models. First, configurations `base$_{\rm H20}$' and `base$_{\rm L20}$' present a similar shape for the scaling relations that set the efficiency of (re)heating gas. These two models present the highest efficiency in massive halos ($V_{\rm max} \gtrsim110\,\rm{km/s}$) compared to the rest of the configurations. Configuration `base$_{\rm H20}$' has the highest efficiency and, consequently, predicts the highest number density of MQs. As discussed in the last subsection, models that favour the production of DSFGs require a lower (re)heating efficiency for high-mass systems, so it is natural for these two configurations to critically underpredict the sub-mm number counts. The SMBH mass functions at $z = 0$ for these configurations show that the SMBHs grew less than the observed; in these configurations, the AGN feedback may be too weak, requiring stronger SN feedback to match the fraction of quiescent galaxies (which is used as a constraint).

On the other hand, the best-fit model of configuration `no HIMF' closely matches (still slightly underpredicting) the observationally-derived SMBH mass function at $z = 0$. For that configuration, an ‘in-between’ scaling relation for (re)heating gas is obtained, yielding reasonably good agreement with the observed sub-mm number counts. These results suggest that the SMBH mass function (and ideally, its evolution) could be a key observable to calibrate galaxy formation models, as it would help in constraining AGN feedback, breaking the degeneracies seen among the main quenching mechanisms (Figure \ref{fig:deg_model}).

\subsection{Limitations and caveats}
Although we found a model that matches the sub-mm number counts reasonably well and simultaneously agrees with the observed lower limits for the number density of massive quiescent galaxies (configuration `no HIMF') using the \citet{henriques20} version of the \texttt{L-Galaxies} SAM, it still presents limitations. For instance, all configurations struggle to capture the massive end of the stellar mass function at high redshift (Figure \ref{fig:smfs}), even for `no low-$z$ SMF, $f_{\rm Q}$', which was specifically designed to match it (using the SMF at $z = 2.8$ as a constraint). This trend is also observed in the \texttt{SHARK} SAM \citep{lagos24}. Moreover, all best-fit models overpredict the cosmic star formation rate density by $\sim0.5\,\rm{dex}$ compared to observational data, except for configuration `base$_{\rm H20}$'. This suggests that the overprediction of the CSFRD is likely due to the \citet{leja20} SMFs, which were used to calibrate these models. Consequently, these results highlight the importance and impact of the observational data used to constrain the models. As mentioned earlier, observational works estimate galaxy properties using a specific method or technique, and these estimates can vary depending on the approach taken. This variability makes the comparison between observations and simulations challenging. A possible solution could be to select observationally-analogous simulated galaxy populations based on forward-modelled observer-frame magnitudes/fluxes, thus avoiding the complex dependence on the techniques used to select galaxies and estimate their properties \citep[see e.g.][]{cochrane_dark}. However, this approach involves other assumptions in the forward-modelling. 

In this work, we assumed that $40\%$ of the metals in the cold gas reservoirs of galaxies are in the form of dust, as the version of \texttt{L-Galaxies} by \citet{henriques20} does not track the evolution of dust. More recently, a detailed model for dust production and destruction has been incorporated into the 2020 version of \texttt{L-Galaxies} \citep{yates24}, which could be employed to refine predictions of sub-mm emission within the simulation. Nevertheless, our assumption is unlikely to significantly affect our results, as adopting a constant dust-to-metal ratio is a reasonable approximation for massive, metal-rich galaxies such as those comprising the SMG population \citep{Remy2014}.

Another crucial finding is the high level of degeneracy (Figure \ref{fig:deg_model}) observed. Although some sets of free parameters yield similar likelihoods, they may represent entirely different treatments of the astrophysical mechanisms. This arises from the hyperparameter space of \texttt{L-Galaxies} (and galaxy formation models in general), which is composed of 15 free parameters. As a result, it is expected that the algorithm will identify multiple reasonable "good" solutions. In this context, robust calibration should be considered as an essential step for galaxy formation models.

\section{Summary} \label{sec:summary}
In this work, we used the Markov Chain Monte Carlo (MCMC) mode of the \texttt{L-Galaxies} semi-analytic model to robustly calibrate its free parameters. Our main goal was to investigate whether this approach can address the long-standing tension between modelled dusty star-forming galaxies (DSFGs; also known as sub-millimetre galaxies, SMGs) and massive quiescent galaxies at high redshifts. To address this, we implemented nine sets of observational constraints, including the widely used stellar mass functions (SMFs), the fraction of quiescent galaxies ($f_{\rm Q}$), and the neutral hydrogen mass function (HIMF), as well as including for the first time the number density of SMGs ($n_{\rm SMG}$). These nine combinations of observational constraints, referred to as configurations (see Table \ref{tab:configs}), produced nine distinct models. Our main predictions and interpretations are as follows:

\begin{enumerate}
\item The SMF is a key observable used to calibrate galaxy formation models. Here, we updated the SMFs used to calibrate the \citet{henriques20} version of \texttt{L-Galaxies} to the \citet{leja20} results. Our predicted SMFs (Figure \ref{fig:smfs}) are consistent with the observational data at low redshift. However, the models struggle to reproduce the massive end at higher redshifts, even when high-redshift SMFs are used as constraints.

\item Similarly, despite incorporating $f_{\rm Q}$ as a function of stellar mass at high redshifts as a constraint, the models fail to reproduce this observable (see Figure \ref{fig:fqs}). When $f_{\rm Q}$ is not used as a constraint at low redshifts, the quiescent population is significantly underrepresented. As expected, configurations excluding $n_{\rm SMG}$ as a constraint exhibit better consistency with the observed $f_{\rm Q}$.

\item Models calibrated with $n_{\rm SMG}$ accurately reproduce the observed $S_{870}$ number counts (Figure \ref{fig:number-counts}). In contrast, models that do not include this constraint critically underpredict this observable, emphasizing the importance of using $n_{\rm SMG}$ in the calibration process, to capture the SMG population.

\item Models that align better with $S_{870}$ number counts tend to underpredict the number density of massive quiescent galaxies at high redshift (Figure \ref{fig:nq_ev}), highlighting the persistent tension between these two populations. However, one configuration — calibrated using SMFs and $f_{\rm Q}$ at $z = 0.4$ and 2.8, and $n_{\rm SMG}$ at $z = 2.8$, while excluding the HIMF (`no HIMF') — achieves reasonable agreement with both populations.

\item Most models predict an elevated cosmic star formation density compared to observational data (Figure \ref{fig:sfrd_z}). This discrepancy appears to stem from the use of \citet{leja20} SMFs for calibration, rather than from the inclusion of $n_{\rm SMG}$.

\item Predicted supermassive black hole (SMBH) mass functions at $z = 0$ (Figure \ref{fig:smbh_mf}) show that most models underpredict the number density compared to observations. This leads to weaker AGN feedback, a critical mechanism for regulating star formation in massive galaxies. The `no HIMF' configuration presents the best agreement with observational data.

\item Analyzing star formation across configurations (Figure \ref{fig:phys_sf}), we found that models calibrated with $n_{\rm SMG}$ exhibit reduced star formation efficiency from secular processes, compensated for by increased merger-induced starbursts. This merger-driven star formation may deplete cold gas, limiting subsequent secular star formation in descendant galaxies.

\item Examining SN feedback (Figure \ref{fig:phys_sn}) reveals that models that better reproduce $f_{\rm Q}$ exhibit higher efficiencies in heating cold gas within massive halos, likely driving their predictions for $f_{\rm Q}$. However, gas ejection efficiencies (outflows) for massive galaxies are similar across models, with significant variations only at the low-mass end. %No clear trends on galaxy properties predictions emerge from these variations.

\item For black hole growth and AGN feedback (Figure \ref{fig:phys_agn}), the fitted models show two distinct trends in SMBH growth via merger-driven cold gas accretion. One trend shows accretion nearly independent of halo mass, resulting in a peaked SMBH mass function at $z = 0$. The other trend exhibits a strong halo mass dependency, producing an SMBH mass function similar to the SMF. The former trend aligns better with observations. Insufficient SMBH masses imply under-powered AGN feedback, except in the `no HIMF' configuration.

\item Finally, we assessed the degeneracy in the `no HIMF' configuration (Figure \ref{fig:deg_model}), which provides a reasonable prediction of sub-millimetre number counts and is consistent with the lower limits of the number density of high-redshift massive quiescent galaxies. Our robust analysis reveals that models with different underlying physics can yield similar likelihood values, owing to significant degeneracies within the hyperparameter space of \texttt{L-Galaxies}. This highlights that the commonly used observational constraints are insufficient to uniquely determine a single preferred physical model. In this context, additional, previously unconsidered constraints — such as SMBH mass functions — may prove valuable in breaking these degeneracies.
\end{enumerate}
Our results provide a comprehensive analysis of calibration outcomes for \texttt{L-Galaxies} and emphasize the importance of robust calibration techniques in exploring the hyperparameter space of galaxy formation models. Additionally, we identify a promising model that represents a step forward in resolving the tension between modelled SMGs and high-redshift massive quiescent populations.

\section*{acknowledgments}
This work was initiated during the CCA Pre-doctoral Program. PA-A thanks the Coordenaç\~o de Aperfeiçoamento de Pessoal de Nível Superior – Brasil (CAPES), for supporting his PhD scholarship (project 88882.332909/2020-01). RKC was funded by support for program \#02321, provided by NASA through a grant from the Space Telescope Science Institute, which is operated by the Association of Universities for Research in Astronomy, Inc., under NASA contract NAS 5-03127. RKC is grateful for support from the Leverhulme Trust via the Leverhulme Early Career Fellowship. The Flatiron Institute is supported by the Simons Foundation. CCH thanks Neal Katz for insightful discussions about semi-analytic models. LSJ acknowledges the support from CNPq (308994/2021-3)  and FAPESP (2011/51680-6). MCV acknowledges the FAPESP for supporting his PhD (2021/06590-0).

%%%%%%%%%%%%%%%%%%%%%%%%%%%%%%%%%%%%%%%%%%%%%%%%%%
\section*{Data Availability}

The full source code for the original L-GALAXIES 2020 version is publicly available via \href{https://github.com/LGalaxiesPublicRelease/LGalaxies2020_PublicRepository}{GitHub}, with installation and running instructions provided on the \href{https://lgalaxiespublicrelease.github.io/}{L-Galaxies website}. Additional simulation data products derived by the authors and presented here can be obtained upon request by sending an email to paraya-araya@usp.br.



\bibliographystyle{mnras}
\bibliography{sample631} % if your bibtex file is called example.bib

%%%%%%%%%%%%%%%%%%%%%%%%%%%%%%%%%%%%%%%%%%%%%%%%%%

%%%%%%%%%%%%%%%%% APPENDICES %%%%%%%%%%%%%%%%%%%%%

\appendix
\section{Observables from the sample of merger trees} \label{sec:sampling_comp}

\begin{figure}
    \centering
    \includegraphics[width=\columnwidth]{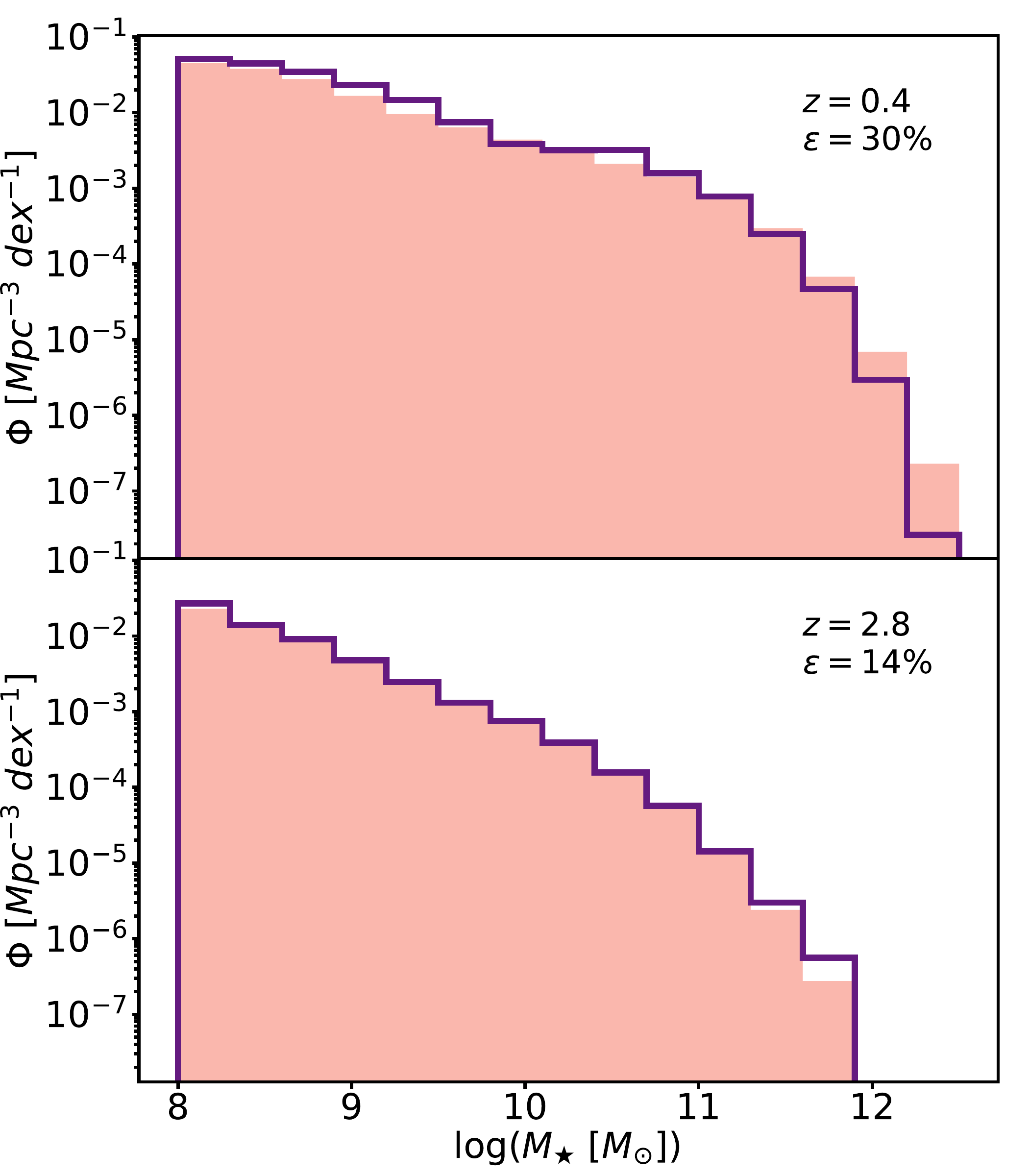}
    \caption{The stellar mass function (SMF) obtained from the fiducial model (orange histogram) compared to that derived from the sample of merger trees (purple solid line) at $z = 0.4$ (top panel) and $z = 2.8$ (bottom panel). The $\epsilon$ value denotes the average relative error between the SMFs (sampled and full volume).}
    \label{fig:sampled_smf}
\end{figure}

\begin{figure}
    \centering
    \includegraphics[width=\columnwidth]{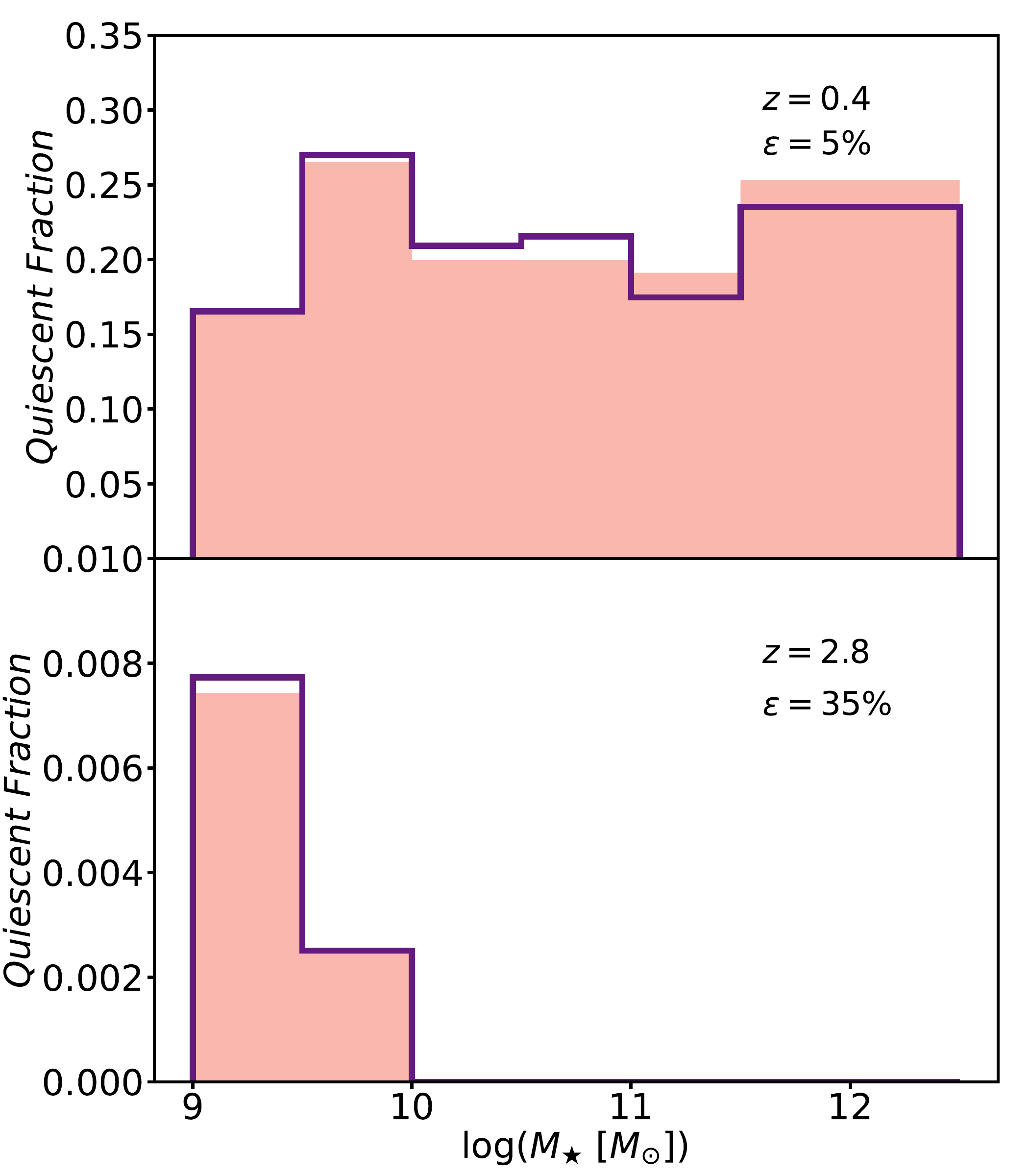}
    \caption{The fraction of quiescent galaxies as a function of stellar mass ($f_{\rm Q}$) obtained from the fiducial model (orange histogram) compared to that derived from the sample of merger trees (purple solid line) at $z = 0.4$ (top panel) and $z = 2.8$ (bottom panel). The $\epsilon$ value denotes the average relative error between both $f_{\rm Q}$ values (sampled and full volume).}
    \label{fig:sampled_fqs}
\end{figure}

In Section \ref{sec:sampling}, we present our method for obtaining a sample of merger trees, the predictions of which must closely represent those of the full simulation volume. The predictions from the sampled merger trees are derived by weighting a given observable according to the fraction of similar haloes not included in the sample. Here, we compare the full-volume and sampled stellar mass functions, as well as the fractions of quiescent galaxies as a function of stellar mass, at $z = 0.4$ and $z = 2.8$, shown in Figure \ref{fig:sampled_smf} and Figure \ref{fig:sampled_fqs}, respectively. To quantify the similarities between the sample and full-volume predictions, we estimate the average relative error, $\epsilon$, as defined by Equation \ref{eq:epsilon}.

\begin{equation} 
\epsilon = \frac{1}{N} \sum\limits_{i=1}^{N} \frac{\left| \Phi_{{\rm sampled},i} - \Phi_{{\rm full},i} \right|}{\Phi_{{\rm full},i}},
\label{eq:epsilon} 
\end{equation}
where $\Phi_{{\rm sampled},i }$ and $\Phi_{{\rm full},i }$ represent a given prediction (e.g. the SMF or $f_{\rm Q}$) in the $i$-th bin — such as a stellar mass bin — from the sample of merger trees and the full volume, respectively. $N$ is the total number of bins.
As can be seen from these figures, the stellar mass function constructed from the sample of merger trees matches that obtained from the full volume well. 

As discussed in Section \ref{sec:sampling}, the merger tree sample was also selected to be representative of the number density of SMGs, another key observable in this work. Although the sampled set comprises only $430$ merger trees out of a total of approximately $20$ million, we obtain comparable predictions for both observables used to calibrate the model. At $z = 2.8$, the fiducial model predicts an SMG number density ($S_{870} \geq 5.2\,\rm{mJy}$) of $n_{\rm SMG} = 9 \times 10^{-6}\,\rm{Mpc^{-3}}$, while the sampled merger tree set yields $n_{\rm SMG} = 8.8 \times 10^{-6}\,\rm{Mpc^{-3}}$.

\section{Best-fitting parameters for each calibration configuration} \label{sec:best-fits}

\begin{figure*}
    \centering
    \includegraphics[width=\textwidth]{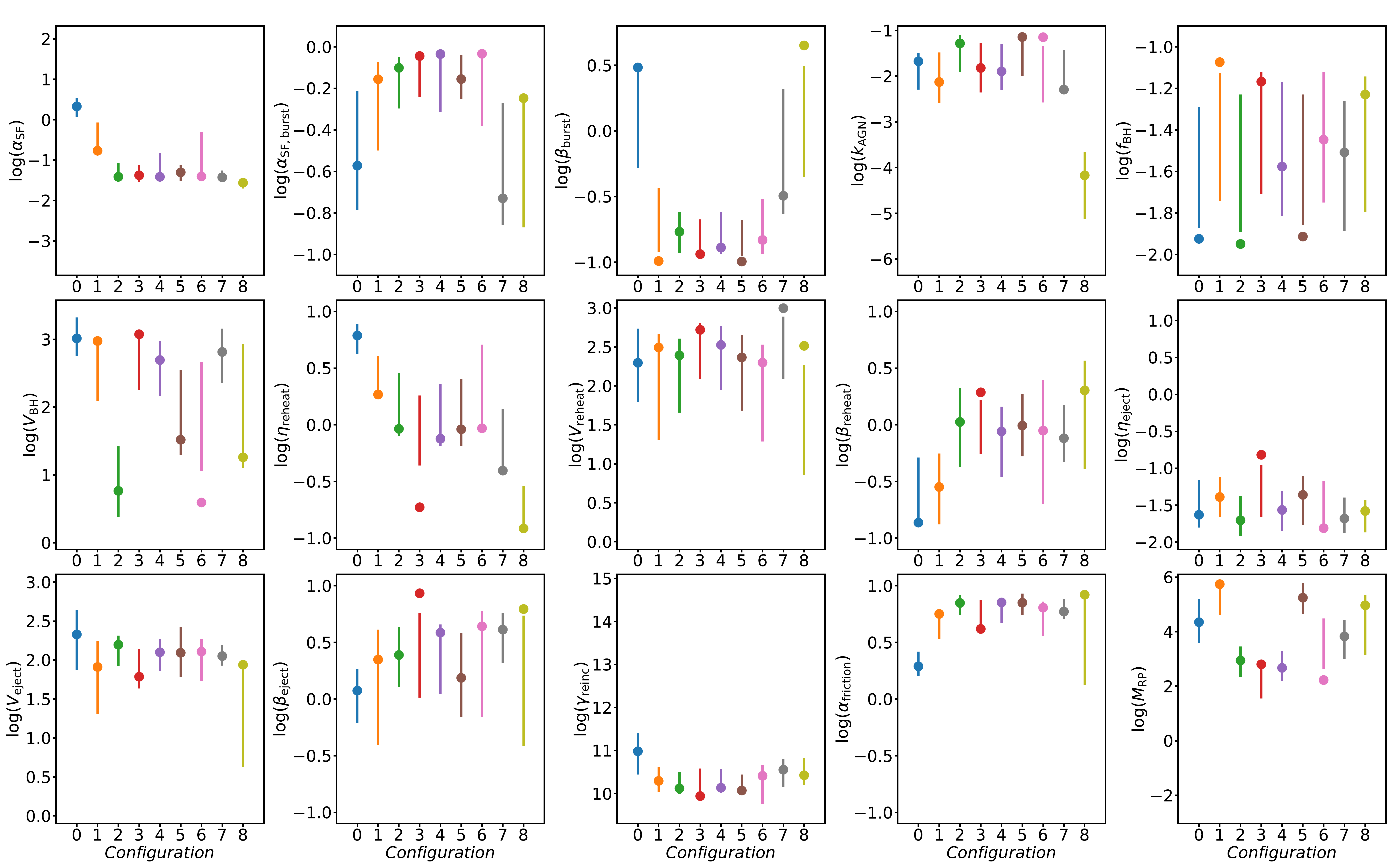}
    \caption{The best-fit parameters obtained for each configuration are listed in Table \ref{tab:configs}. The error bars represent the $16^{\rm{th}}$ and $84^{\rm{th}}$ percentiles of the final $2,000$ MCMC runs across $96$ chains. The parameters $\alpha_{\rm SF}$ (first panel in the top row) and $k_{\rm AGN}$ (fourth panel in the top row) are shown in the top panel of Figure \ref{fig:phys_sf} and the bottom panel of Figure \ref{fig:phys_agn}, respectively.}
    \label{fig:model_comp}
\end{figure*}

We present the best-fitting values for the 15 free parameters across the nine configurations listed in Table \ref{tab:configs} in Figure \ref{fig:model_comp}. The error bars in Figure \ref{fig:model_comp} correspond to the $16^{\rm{th}}$ and $84^{\rm{th}}$ percentiles from the final 2,000 MCMC samples across 96 chains (see Section \ref{sec:run}).

The parameters $\alpha_{\rm SF}$, $\alpha_{\rm SF, burst}$, and $\beta_{\rm SF, burst}$ are associated with secular star formation (as shown in the top panel of Figure \ref{fig:phys_sf}) and merger-induced starbursts (free parameters of Equation \ref{eq:sf_burst}; bottom panel of Figure \ref{fig:phys_sf}), respectively. The AGN efficiency parameter, $k_{\rm AGN}$, is the same as presented in the bottom panel of Figure \ref{fig:phys_agn}, while $f_{\rm BH}$ and $V_{\rm BH}$ are the free parameters governing the black hole growth scaling relation (Equation \ref{eq:quasar_mode}; top panel of Figure \ref{fig:phys_agn}).

The parameters $\eta_{\rm reheat}$, $V_{\rm reheat}$, and $\beta_{\rm reheat}$ define the scaling relation in Equation \ref{eq:sn_feedback} describing gas (re)heating due to supernova feedback (Figure \ref{fig:phys_sn}, top panel). Similarly, the parameters $\eta_{\rm eject}$, $V_{\rm eject}$, and $\beta_{\rm eject}$, which govern the ejection of gas from the hot gas atmosphere, share the same scaling form as the (re)heating process (Figure \ref{fig:phys_sn}; bottom panel).

The remaining parameters, not discussed in detail in this paper, include $\gamma_{\rm reinc}$, which sets the reincorporation timescale of ejected gas; $\alpha_{\rm friction}$, a correction factor for dynamical friction based on the \citet{binney87} formula; and finally, $M_{\rm RP}$, the halo mass threshold above which ram pressure stripping is considered.

%%%%%%%%%%%%%%%%%%%%%%%%%%%%%%%%%%%%%%%%%%%%%%%%%%

% Don't change these lines
\bsp	% typesetting comment
\label{lastpage}
\end{document}